\documentclass[twocolumn,times]{aastex631}

\hypersetup{linkcolor=green,citecolor=magenta,filecolor=cyan,urlcolor=red}
\usepackage{rotating}
\usepackage{mathrsfs}
\usepackage{wrapfig}
\usepackage{graphicx}
\usepackage{xcolor}
\usepackage{hyperref}

\begin{document}

\title{A Relativistic Jet in the Radio Quiet AGN Mrk 110}

\correspondingauthor{Ailing Wang}
\email{wangal@ihep.ac.cn}
\email{wal@shao.ac.cn}
\correspondingauthor{Tao An}
\email{antao@shao.ac.cn}

\author[0000-0002-7351-5801]{Ailing Wang}
\affiliation{Key Laboratory of Particle Astrophysics, Institute of High Energy Physics, Chinese Academy of Sciences, Beijing 100049, China}
\affiliation{Shanghai Astronomical Observatory, Chinese Academy of Sciences, 80 Nandan Road, Shanghai 200030, China}
\affiliation{University of Chinese Academy of Sciences, 19A Yuquan Road, Beijing 100049, China;}
\author[0000-0003-4341-0029]{Tao An}
\affiliation{Shanghai Astronomical Observatory, Chinese Academy of Sciences, 80 Nandan Road, Shanghai 200030, China}
\affiliation{University of Chinese Academy of Sciences, 19A Yuquan Road, Beijing 100049, China;}
\affiliation{State Key Laboratory of Radio Astronomy and Technology, A20 Datun Road, Chaoyang District, Beijing, P. R. China}
\author[0000-0002-0093-4917]{Kenneth I. Kellermann}
\affiliation{National Radio Astronomy Observatory, 520 Edgemont Rd., Charlottesville, VA 22903, USA}
\author[0000-0001-7584-6236]{Hua Feng}
\affiliation{Key Laboratory of Particle Astrophysics, Institute of High Energy Physics, Chinese Academy of Sciences, Beijing 100049, China}
\author[0000-0002-1727-1224]{Emmanuel K. Bempong-Manful}
\affiliation{Jodrell Bank Centre for Astrophysics, Department of Physics and Astronomy, The University of Manchester, Manchester M13 9PL, UK}
\affiliation{School of Physics, University of Bristol, Tyndall Avenue, Bristol BS8 1TL, UK}
\author[0000-0001-9404-2612]{Roland Timmerman}
\affiliation{Centre for Extragalactic Astronomy, Department of Physics, Durham University, Durham, DH1 3LE, UK}
\affiliation{Institute for Computational Cosmology, Department of Physics, Durham University, South Road, Durham DH1 3LE, UK}
\author[0000-0003-0181-7656]{Shaoguang Guo}
\affiliation{Shanghai Astronomical Observatory, Chinese Academy of Sciences, 80 Nandan Road, Shanghai 200030, China}
\affiliation{University of Chinese Academy of Sciences, 19A Yuquan Road, Beijing 100049, China;}
\affiliation{State Key Laboratory of Radio Astronomy and Technology, A20 Datun Road, Chaoyang District, Beijing, P. R. China}

\begin{abstract}

We report the discovery of a relativistic jet in Mrk~110, a narrow-line Seyfert 1 galaxy historically classified as a radio-quiet active galactic nucleus (AGN). Very Long Baseline Interferometry (VLBI) observations reveal intermittent jet activity during 2015--2016 and 2022--2024, with proper motion measurements yielding superluminal velocities of $\sim3.6\pm0.6\,c$ and $\sim2.1\pm0.2\,c$, respectively. The recent jet component decelerates to $\sim 1.5\pm0.2\,c$ at a projected distance of 1.1 parsec from the core, 
coinciding with the transition zone between broad-line and narrow-line regions. This deceleration accompanies  dramatic spectral evolution from steep (the spectral index $\alpha \approx -0.63 \pm 0.04$) to inverted ($\alpha \approx +0.69 \pm 0.10$) as the 7.6 GHz flux density more than doubled. 
These episodic jet ejections and their  evolutionary pattern  match 
theoretical predictions from magnetically arrested disk (MAD) models for temporary jet formation in systems with Mrk 110's physical parameters on timescales of months to years. The observed jet deceleration distance matches expectations for relativistic outflows interacting with the circumnuclear environment. These findings demonstrate that the traditional radio-loud/quiet AGN dichotomy may reflect time-averaged states rather than intrinsic capabilities, suggesting that jets may form across the AGN population but become observable only during specific accretion phases when MAD conditions are temporarily established.
Mrk 110 serves as a critical ``missing link'' between radio-loud and radio-quiet AGN, providing insight into jet formation mechanisms, environmental interactions, and physical processes that unify various AGN classifications.
\end{abstract}


\section{Introduction} \label{sec:intro}

The radio properties of active galactic nuclei (AGN) have traditionally been used as a fundamental classification criterion, dividing the population into radio-loud and radio-quiet classes based on the ratio of radio to optical flux densities \citep{1989AJ.....98.1195K,1994AJ....108.1163K}. This dichotomy has long been interpreted as reflecting intrinsic differences in jet production efficiency, with radio-loud AGN capable of producing powerful relativistic jets while radio-quiet sources presumably lack significant jet activity. 
However, theoretical advances and observational studies have begun revealing additional complexity within this bimodal paradigm, suggesting that jet formation capabilities might be more widespread but episodic in nature \citep{2009ApJ...698..840C, 2013ApJ...764L..24S}, requiring sensitive observations of traditionally radio-quiet sources to detect transient relativistic outflows.

Narrow-Line Seyfert 1 galaxies (NLS1s) represent ideal laboratories for testing these competing models. Characterized by narrow permitted lines (FWHM(H$\beta$) $< 2000$~km~s$^{-1}$), strong Fe~II emission, and relatively weak [O~III] emission \citep{1985ApJ...297..166O,1992ApJS...80..109B}, NLS1s typically display radio-quiet properties despite their high accretion rates.
Mrk~110 ($z = 0.035$\footnote{At the angular distance of 143.8 Mpc, 1 mas corresponds to 0.7 parsec, and 1 mas yr$^{-1}$ proper motion corresponds to $2.35\, c$. Its proximity enables unprecedented physical resolution to study the jet structure and kinematics.}) stands as a particularly intriguing case, with reverberation mapping measurements revealing a black hole mass ($M_{\rm BH} \sim 2 \times 10^{7}~M_{\odot}$) and high accretion rate ($\sim0.40~L_{\rm Edd}$) \citep{2004ApJ...613..682P,
2003A&A...407..461K, 2019ApJ...886...42D,2022MNRAS.512L..33V} that place it at the upper end of the NLS1 mass distribution. X-ray observations have revealed complex spectral and temporal behaviors, including significant soft excess emission and rapid variability characteristic of NLS1s \citep{2006ApJ...651L..13D, 2021A&A...654A..89P,2021MNRAS.504.4337V,2022MNRAS.510..718P}.

In the radio regime, Mrk~110 has been classified as radio-quiet based on conventional criteria \citep[$R \sim 1.49$ \footnote{We adopt the classical definition of radio-loudness, using the radio-loudness $\mathrm{ R=S_{5 GHz}/S_{4000\AA}}$ (ratio of radio to optical flux densities), with $R<10$ defining radio-quiet AGN.},][]{1989AJ.....98.1195K}. The sensitivity and resolution limitations of earlier radio observations have left significant questions regarding the potential presence of low-power jets or other radio structures. Recent Very Long {Baseline} Array (VLBA) observations reveal an unresolved, variable core (0.71 -- 1.21 mJy at C band) confined to an extremely compact region ($\leq 180$ Schwarzschild radii) with high brightness temperature ($6.0 - 15.7 \times 10^7$ K) \citep{2023MNRAS.518...39W,2023MNRAS.525.6064W,2022MNRAS.510..718P}, coexisting with extended kiloparsec-scale structures recording historical AGN jet activity over Myr timescales  \citep{1994AJ....108.1163K, 1998MNRAS.297..366K, 2022A&A...658A..12J}. This combined compact and extended morphology alongside radio-quiet classification and variable core establishes Mrk 110 as an exceptional laboratory for studying weak radio emission mechanisms in AGN, addressing fundamental questions about jet formation conditions and feedback in traditionally non-relativistic systems \citep{2012SSRv..169...27P, 2016MNRAS.461..967M}, the role of low-power jets in AGN feedback \citep{2022MNRAS.512.1608G, 2024A&A...685A.122U}, and relationships between intermittent radio activity and accretion processes \citep{2009ApJ...698..840C,2013ApJ...764L..24S, 2023ApJ...950...31K}. Similar VLBI signatures have been observed in other radio-quiet AGNs, where proper motion studies reveal moderate relativistic jet components (e.g., PG 1351+640: \citealt{2023MNRAS.523L..30W}; III Zw 2: \citealt{2000A&A...357L..45B}), demonstrating that compact jets can form and evolve even in low-power systems.

In this Letter, we present new Very Long Baseline Interferometry (VLBI) observations of Mrk~110, designed to probe its radio structure with high resolution and sensitivity. 
Our investigation aims to test whether the conventional radio-loud/quiet dichotomy accurately describes the underlying physics of AGN jet formation and propagation.

\section{Observations and Data Reduction}

We performed an extensive multi-frequency VLBI observation campaign of Mrk 110 between 2015 and 2024, utilizing both the VLBA and European VLBI Network (EVN) at 1.6, 4.7, 4.9, and 7.6 GHz. This strategic observational approach was designed to optimize sensitivity to weak, compact radio emission while providing sufficient temporal and spectral coverage to characterize the source's structural and spectral evolution across multiple timescales. We present here previously unpublished observations from 2023-2024, which complement our earlier exploratory observations initiated in 2015 \citep{2023MNRAS.518...39W} and subsequent systematic monitoring during 2021--2022 \citep{2023MNRAS.525.6064W}. Following the detection of enhanced radio activity, we increased our observational cadence throughout 2023--2024. The complete observational parameters and scheduling details are provided in Appendix \ref{app:VLBI}.  Data calibration and imaging followed standard VLBI procedures implemented through the Astronomical Image Processing System (\textsc{aips}) and \textsc{Difmap}  software packages. 
Additionally, we re-analyzed the archival VLBA data from \citet{2022MNRAS.510..718P} observed in 2015 August -- 2016 May. 
The final images achieved typical noise levels of 19--40 $\mu$Jy beam$^{-1}$, and resolutions of $2.6 \times 0.9$ mas$^2$ at 7.6 GHz and $4.7 \times 2.0$ mas$^2$ at 4.7 GHz.
To extend our analysis to lower frequencies and larger spatial scales, we also processed archival International LOFAR observations of Mrk 110 at 144 MHz (see Appendix \ref{app:lofar} for details).

\section{Results}

\subsection{Parsec-Scale Structure and Kinematics} \label{sec:3.1}

VLBI images consistently show Mrk 110 as a single compact component (Fig. \ref{fig1}-a and Fig. \ref{fig:VLBI_image}) without extended emission on parsec scales. Multi-epoch observations reveal systematic northwestern (along position angle of $\sim -29\degr \pm 12\degr$) displacement of the peak position by $\sim 1.6$ mas  ($\sim$1.1 pc) from 2021 December to 2024 February (Figures \ref{fig1}-c). Multiple lines of evidence rule out instrumental or calibration effects of this displacement: (1) Position shift occurs consistently across two frequencies; (2) The phase reference calibrator remained stable with positional consistency $<0.05$ mas; (3) The array configuration remained consistently (similar restoring beams seen in the VLBA images in Fig. \ref{fig:VLBI_image});  and (4) The motion direction aligns with the kpc-scale emission detected by international LOFAR (Fig. \ref{fig1}-b).

Despite appearing as a single component, this morphology is consistent with  jet evolution in radio-quiet AGN. Similar characteristics have been observed in relativistic jets from tidal disruption events (TDEs) such as Swift J1644+57 \citep{2011Natur.476..425Z, 2012ApJ...748...36B}, which maintain compact appearance at VLBI scales while exhibiting dramatic spectral changes during their initial stages. In these cases, peak position shifts and spectral evolution serve as reliable jet indicators before any dual-component structure becomes resolvable.

Between 2021 December and 2023 August, the spectral index transitioned from steep  ($ \alpha \approx -0.63 $) to inverted ($\alpha = +0.15$), suggesting the emergence of a new self-absorbed jet component. The positional shift implies a minimum apparent speed of $\beta_{\rm app} > 2.1$ (in unit of the speed of light, $c$) at 7.6 GHz assuming jet ejection in early 2022 (Fig. \ref{fig:pm_7GHz}), constraining jet parameters to $\Gamma_{\rm min} \geq 2.3$ and $\theta \leq 25.5^\circ$. 
These values are conservative limits, as any later ejection time would require more extreme values. 

The jet evolution reveals a dramatic change after 2022 (Fig.~ \ref{fig1}-c and \ref{fig1}-d). Analysis of 7.6-GHz observations between 2023 August and 2024 February yields a measured proper motion of $0.6 \pm 0.1$ mas~year$^{-1}$, corresponding to an apparent velocity $\beta_{\text{app}} = 1.5 \pm 0.2$ (Fig. \ref{fig:pm_7GHz}). For this later phase, relativistic beaming theory constrains the jet parameters: minimum Lorentz factor $\Gamma_{\text{min}} = 1.8$ and viewing angle $\theta \leq 33.7^\circ$.

The 2015--2016 data, though limited to a shorter monitoring timeline, clearly shows evidence of an earlier jet ejection episode with significant positional shift in the northwest direction, yielding an apparent velocity of $\beta_{\rm app} = 3.6 \pm 0.6$ (Fig. \ref{fig1}-c), remarkably consistent with the 2021 December--2024 February measurements, confirming the relativistic nature of the outflow.

The observed jet deceleration from $\beta_{\rm app} = 2.1 \pm 0.2$ to $\beta_{\rm app} = 1.5 \pm 0.2$ occurs at approximately 1.1 parsec (projected) from the AGN core. Accounting for a likely viewing angle of $10\degr-20\degr$, the deprojected distance would be ~3.2--6.4 pc. 
While AGN structure scales vary significantly across the population \citep{2015ARA&A..53..365N}, Mrk 110's specific parameters allow us to constrain its nuclear geometry. Reverberation mapping results \citep{2003A&A...407..461K,2004ApJ...613..682P} indicate a broad line region (BLR) size of 0.08 pc for Mrk 110, placing our observed deceleration zone well beyond the BLR but coincident with the expected NLR/torus transition region. The significant deceleration observed is consistent with recent hydrodynamical simulations \citep{2018MNRAS.479.5544M} showing that intrinsically low-power jets  like that in Mrk 110 experience substantial slowing through mass entrainment when propagating through dense environments. 
The interaction appears more pronounced than in typical radio-loud AGN, consistent with NLS1 galaxies having denser nuclear environments due to their higher accretion rates \citep{2015A&A...575A..13F,2020A&A...636A..64B}. Our observation provides a rare glimpse into how environmental factors regulate jet propagation in radio-quiet AGN. Our observation provides a rare glimpse into how environmental factors regulate jet propagation in radio-quiet AGN.

\subsection{Spectral and Flux Density Evolution} \label{sec:3.2}

The radio spectrum and flux density of Mrk~110 underwent remarkable changes during our monitoring period (Figure \ref{fig1}-e). 
Between 2015--2016, the source exhibited a typical optically thin synchrotron spectrum ($\alpha \approx -0.76$ to $-0.55$) that showed low-level variability across epochs (also see \citealt{2022MNRAS.510..718P}). 

A dramatic transformation occurred in early 2023, when the spectrum became inverted ($\alpha = +0.15$) primarily due to enhanced 7.6-GHz emission while the 4.7-GHz flux density remained relatively stable. By February 2024, the spectral index increased further ($\alpha = +0.69 \pm 0.10$), suggesting ongoing evolution in the physical properties of the radio-emitting region.  This spectral transition from optically thin to optically thick regimes provides evidence for the emergence of a nascent, self-absorbed jet component.

The transition from steep to inverted spectrum indicates the emergence of a new, self-absorbed jet component that becomes dominant at high frequencies. The pre-existing steep-spectrum emission from the compact nuclear region likely remains at similar levels, but becomes subdominant compared to the new jet component at 7.6 GHz due to the jet's self-absorbed synchrotron emission peaking at higher frequencies. Unlike classical blazars where cores are typically flat-spectrum due to persistent jet self-absorption, the nuclear emission in radio-quiet AGN like Mrk 110 often exhibits steep spectra due to the mixture of optically thin synchrotron emission from various processes. By 2024, the \textbf{nascent} jet component contributes approximately (70-80)\% of the total flux at 7.6 GHz. This frequency-dependent dominance explains why the 7.6 GHz flux density more than doubled while the 4.7 GHz flux showed only modest increase.
Similar spectral transformations were observed in other radio-intermediate AGN like III Zw 2 \citep{1999ApJ...514L..17F}, reinforcing our jet emergence interpretation.

This distinct spectral signature provides quantitative evidence nascent jet evolution. Our measurements show asymmetric flux changes across frequencies: at 4.7 GHz, peak flux density increased moderately from $0.95 \pm 0.04$ to $1.18 \pm 0.02$ mJy beam$^{-1}$ with stable integrated flux density, 
while at 7.6 GHz, both peak and integrated flux densities rose significantly ($0.69 \pm 0.04$ to $1.48 \pm 0.03$ mJy beam$^{-1}$).  Such frequency-dependent amplification is a defining characteristic of emerging relativistic jets transitioning from optically thick to thin regimes at progressively lower frequencies.

The VLBI data reveal that Mrk 110 experiences two episodic jet activities in 2015--2016 and in 2021-2024. 
The 2021--2024 event, captured comprehensively in our monitoring campaign, demonstrates a dramatic spectral evolution from steep ($\alpha \approx -0.63\pm0.04$) to flat ($\alpha \approx +0.15\pm0.10$) to inverted ($\alpha \approx  +0.69\pm0.10$). This progression precisely traces the emergence and evolution of a new jet component detected in our kinematic analysis (Section \ref{sec:3.1}). Unlike the later event, we did not capture any spectral transformation during the 2015--2016 period, suggesting that our observations likely sampled only the optically thin phase before the emergence of the optically thick component. The gap between 2016 and 2021 prevents us from determining whether additional ejections occurred during this interval. This pattern of intermittent jet formation with distinct spectral signatures is significant in a traditionally radio-quiet NLS1, suggesting that episodic jet production may be more common in these sources than previously recognized.  

The observed spectral progression in Mrk 110 quantitatively matches the theoretical models of evolving self-absorbed synchrotron emission in nascent jets \citep{1985ApJ...298..114M}, with the inverted spectrum ($\alpha \approx 0.7$) precisely corresponding to the canonical value for optically thick synchrotron emission from a homogeneous source ($\alpha \approx 0.5-1.0$). Alternative explanations for the observed spectral evolution (including free-free absorption variability, interstellar scintillation, and changing accretion states) would produce more erratic spectral variations or fail to explain the simultaneous structural displacement. The tight temporal correlation between spectral changes and kinematic evolution strongly favors jet emergence.

\subsection{Kpc-Scale Morphology}

\begin{figure*}
\centering
\includegraphics[width=0.4\linewidth]{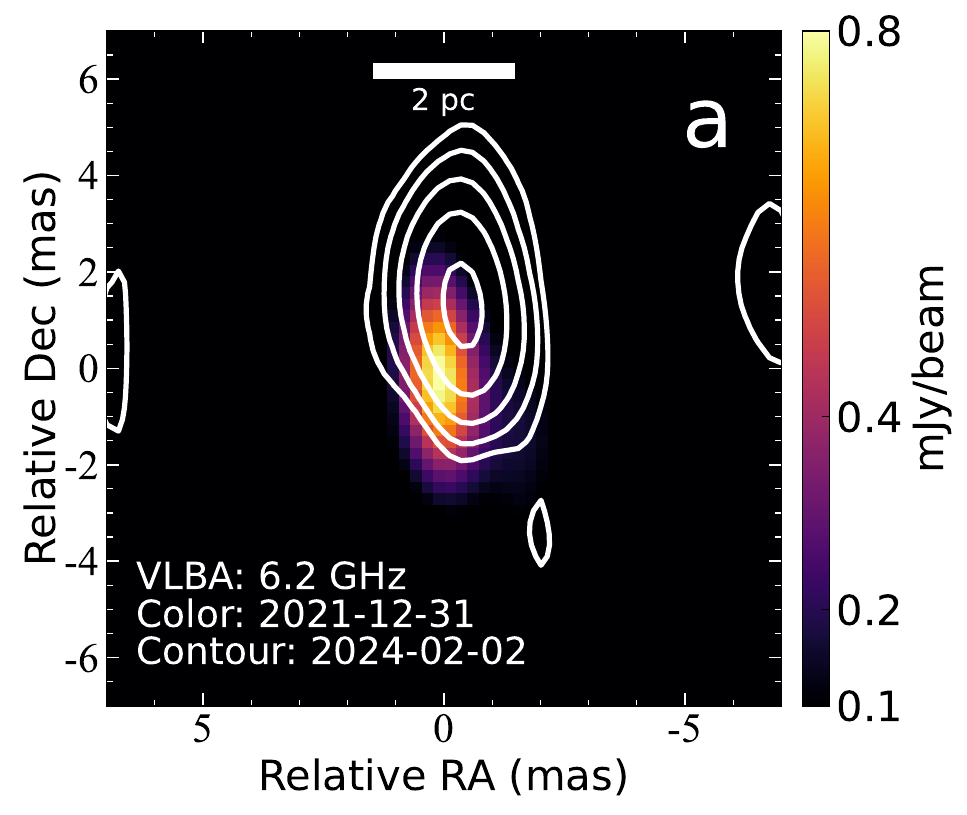} \hspace{2mm}
\includegraphics[width=0.45\linewidth]{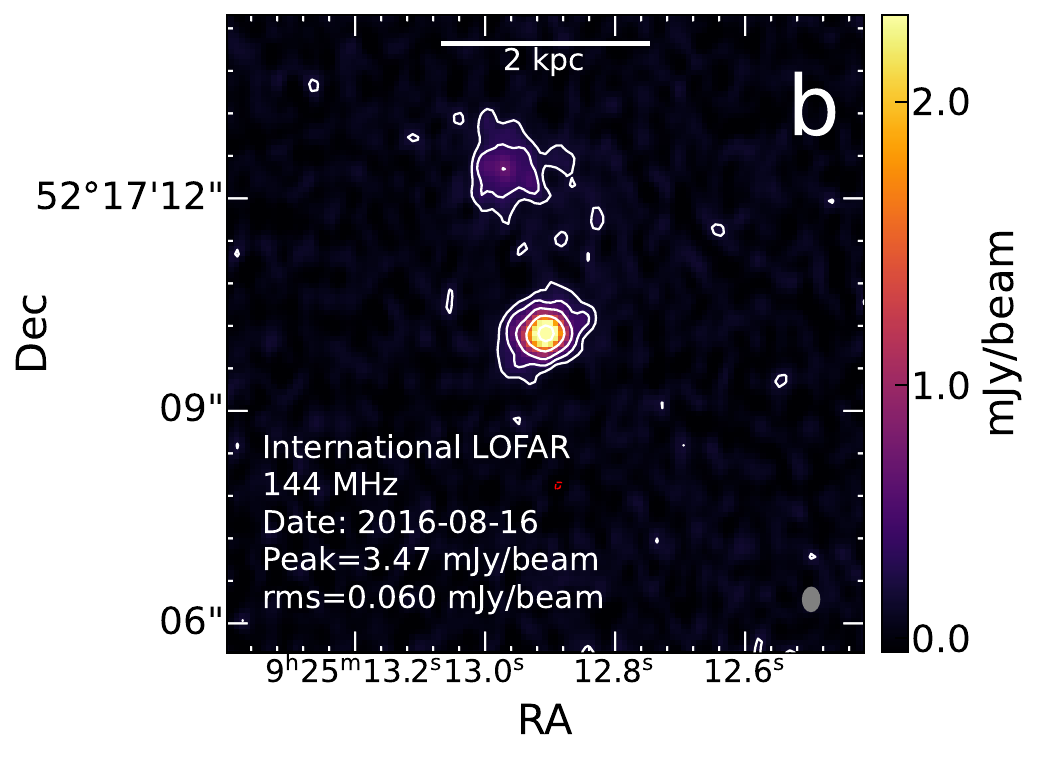} \\    
\includegraphics[width=0.475\linewidth]{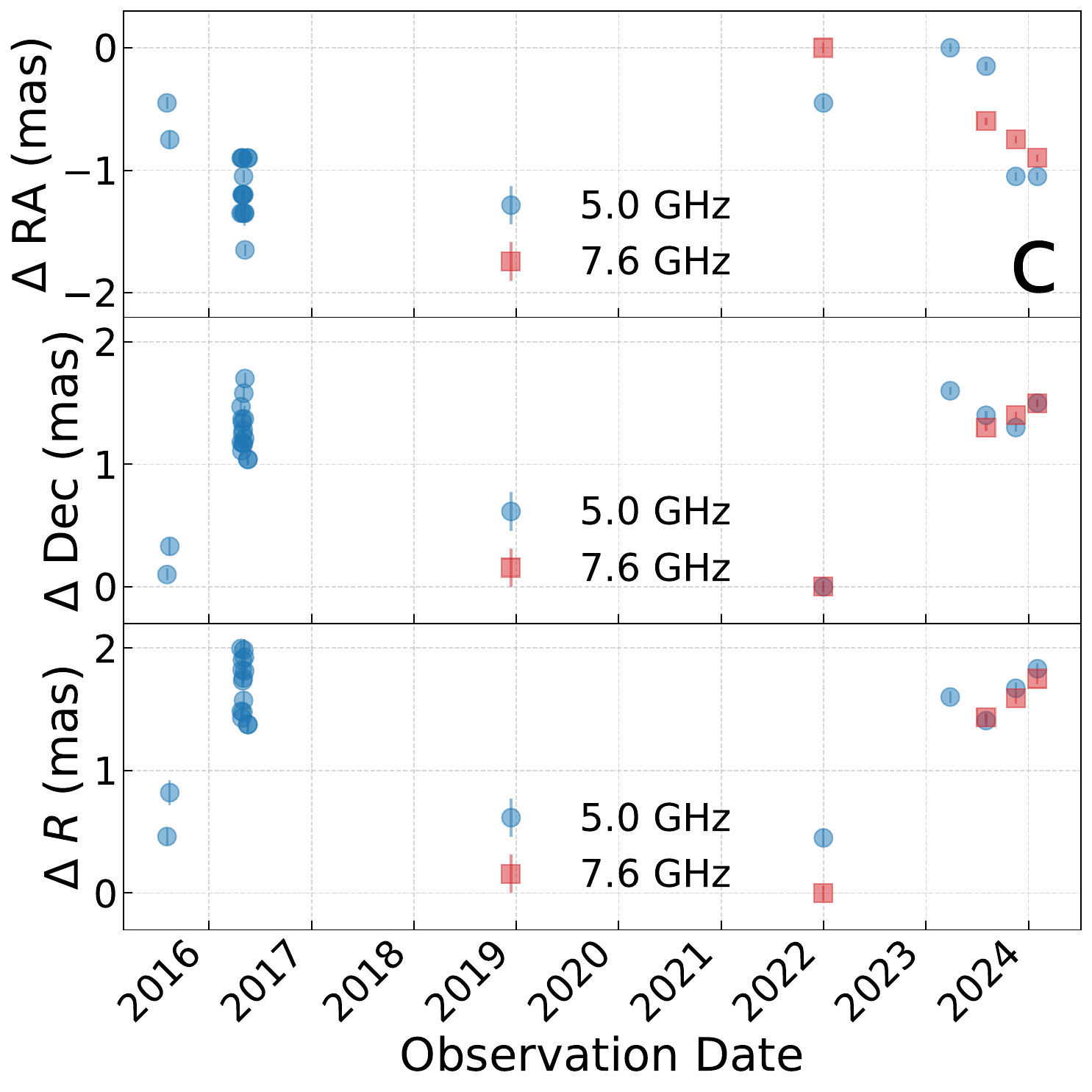} \hspace{2mm}
\includegraphics[width=0.5\linewidth]{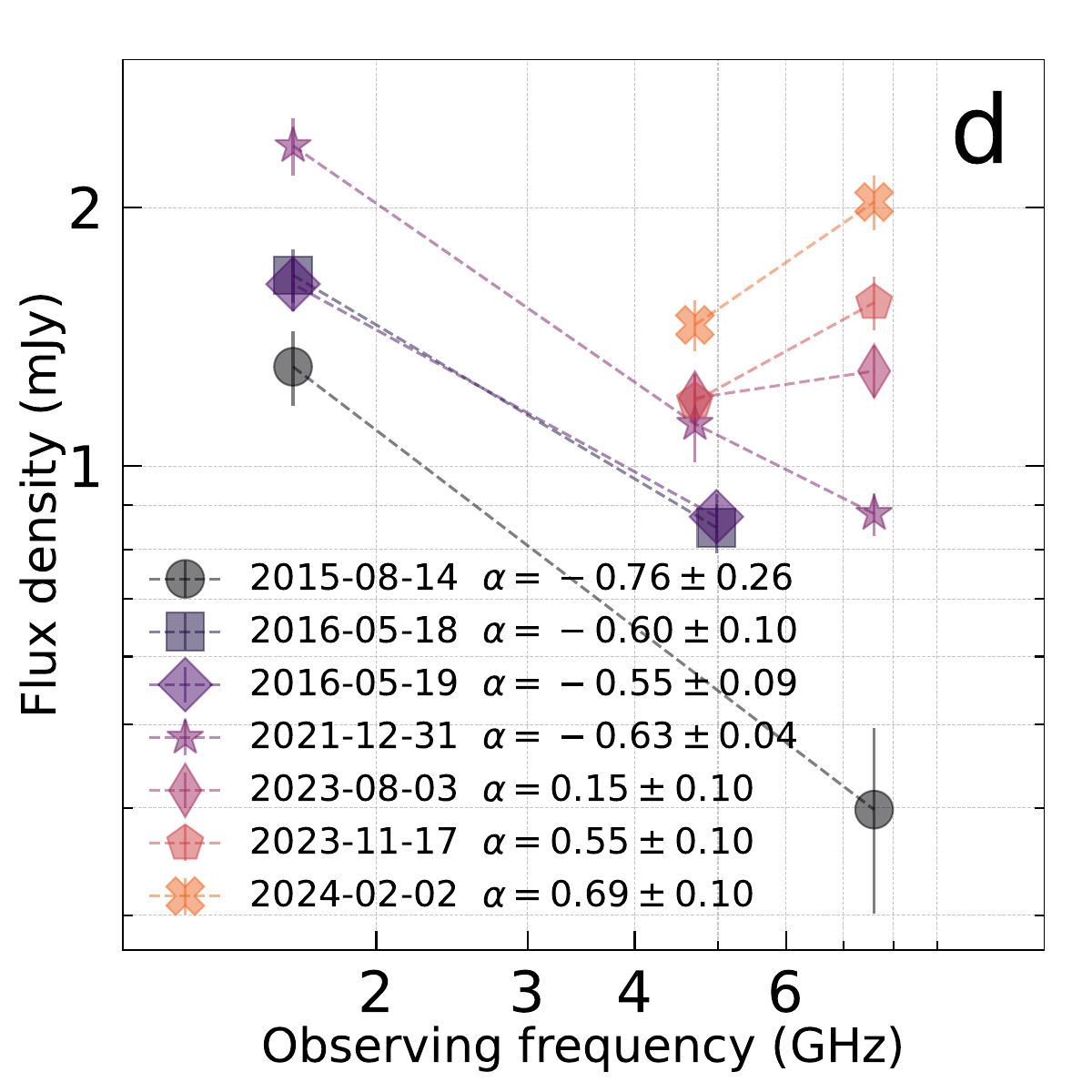}
\caption{Radio observations of Mrk~110. 
(a): VLBA 6.2 GHz images of Mrk 110 from 2021 Dec 31 (color scale) and 2024 Feb 2 (contours), showing core shift of $\sim 1.6$ mas. 
The image center is set at the 2021 December peak position (RA = 09:25:12.84781, Dec = +52:17:10.3862). 
(b): International LOFAR image at 144 MHz observed on 2016 August 16, revealing $\sim1.7$ kpc northern extension aligned with the parsec-scale jet. 
(c): Peak position evolution.
(d): Change of the radio spectrum from 2015 to 2024, showing transformation from a steep ($\alpha = -0.76$) to inverted ($\alpha = +0.69$) spectral index. Historical data (2015--2016) are from \citet{2022MNRAS.510..718P}; our recent measurements are detailed in Table~\ref{tab:obs-info}. Dashed lines indicate power-law fits.
}
\label{fig1}
\end{figure*}

International LOFAR observations at 144 MHz (Fig. \ref{fig1}-b) reveal two distinct components in Mrk 110: a compact feature coinciding with the galactic nucleus and a weak extended feature $\sim1.7$ kpc north. This northern feature, previously detected at higher frequencies  \citep{1994AJ....108.1163K, 1998MNRAS.297..366K, 2022A&A...658A..12J}, is clearly resolved in our international LOFAR data, which combines sub-arcsecond resolution with sensitivity to aged synchrotron emission. The diffuse morphology and steep spectrum of this component resemble classical radio lobes at significantly lower luminosities, consistent with Mrk 110's radio-quiet classification. 

The alignment between this extended emission and our VLBI-detected parsec-scale jet demonstrates long-term directional stability in Mrk 110's outflow axis across widely different timescales. This persistent orientation over Myr periods suggests the central engine repeatedly launches jets along the same preferred axis despite intermittent activity. 
Together, the kpc-scale structure and pc-scale VLBI jet provide complementary evidence of Mrk 110's past and current relativistic activity, offering crucial insight into energy transport mechanisms across multiple scales in radio-quiet AGN.

\section{Discussion}

The discovery of relativistic jets in Mrk 110 enriches our understanding of jet physics in traditionally radio-quiet AGN. Our observations reveal a complex reality where jet formation exists along a continuum that includes diverse manifestations within the established radio-loud/quiet framework, with important implications for both AGN classification and evolution.

\subsection{Jet formation in Radio-Quiet AGN}
\label{sec:Jet Formation}

The detection of relativistic motion in Mrk 110 significantly transforms our understanding of jet physics in AGN classified as radio-quiet. While our observational results establish the presence of superluminal motion and spectral evolution characteristic of jet formation, the theoretical implications extend far beyond these measurements alone.

We interpret the spectral evolution as resulting from the emergence of a new self-absorbed jet component whose emission becomes dominant at high frequencies, while the pre-existing core-region emission (characterized by steep spectrum) remains relatively stable but becomes subdominant at 7.6 GHz. The modest increase at 4.7 GHz reflects the jet's self-absorbed nature at this frequency, while the dramatic increase at 7.6 GHz indicates the jet's spectral turnover occurs at higher frequencies. This observational signature is similar to that of relativistic jets in TDEs \citep{2018Sci...361..482M, 2021ApJ...908..125C, 2022ApJ...927...74M}, where abrupt jet formation produces dramatic spectral hardening and enhanced high-frequency emission before structural complexity becomes apparent. This frequency-dependent jet dominance explains the observed flux density variation and spectral evolution without requiring resolved structural changes in our images.

The physical mechanism behind Mrk 110's jet activity likely involves a combination of key factors that our observations can partially constrain: (1) high accretion rate ($0.58 < \log \dot{M} < 0.92$, \citealt{2019ApJ...886...42D}), facilitating efficient magnetic flux amplification through disk turbulence \citep{2016MNRAS.457..857S, 2024MNRAS.532.1522J}; (2) high black hole spin  ($a_* \approx 1$, \citealt{2024A&A...681A..40P}), providing the rotational energy reservoir needed for relativistic jet launching via the Blandford-Znajek process when coupled with a magnetically arrested disk \citep{2011MNRAS.418L..79T}; and (3) episodic magnetic flux accumulation and dissipation cycles \citep{2012MNRAS.423.3083M}, evidenced by our detection of discrete jet ejection episodes. These observations are consistent with magnetically-driven jet launching processes like the Blandford-Znajek mechanism, though direct measurements of magnetic field configurations remain a target for future investigations.

This paradigm builds upon the established radio-loud/radio-quiet classification by revealing nuances within this framework. Instead, our findings suggest a continuous spectrum of jet activity, where even nominally radio-quiet systems harbor the capacity for transient relativistic outflows when conditions temporarily align. 
General Relativistic Magnetohydrodynamics (GRMHD) simulations support this view, demonstrating that magnetically arrested disk (MAD) states, necessary for efficient jet production, can form sporadically even in systems that appear radio-quiet in time-averaged observations \citep{2023arXiv231100432C, 2024arXiv240618061M}.

We have carefully considered alternative explanations for our observations. Pure opacity variations in synchrotron self-absorption could potentially explain spectral changes but cannot account for the systematic positional shift ($\sim$1.6 mas) without invoking bulk motion. Interstellar scintillation at Mrk 110's Galactic latitude ($b\sim44^\circ$) would produce a modulation index considerably below our observed flux variations ($>100\%$), and cannot generate directional positional shifts. Accretion state transitions might alter spectral properties but occur within stationary emission regions, failing to explain the measured superluminal motion ($\beta_{\rm app} = 1.5$--$3.6$). Even a combination of these mechanisms cannot simultaneously account for the three key phenomena we observe: superluminal positional shifts, steep-to-inverted spectral evolution, and frequency-dependent flux density increases. The relativistic jet interpretation uniquely explains all observed characteristics through a single coherent physical mechanism supported by established theoretical models.

The episodic pattern observed in Mrk 110 provides a crucial missing link in understanding AGN unification schemes.
For Mrk 110's specific parameters, recent theoretical models predict magnetic flux accumulation reaches critical thresholds for jet launching. The magnetohydrodynamic timescale for flux accumulation in systems with Mrk 110's parameters is predicted to be $t_{\rm MAD} \sim 10^5 r_g/c \approx 10^7 s$, 
where $r_g = GM_{\rm BH}/c^2 \approx 3\times10^{12}$ cm is the gravitational radius
at the innermost stable circular orbit \citep{2012MNRAS.423.3083M},
corresponding to approximately 4 months. State-of-the-art GRMHD simulations \citep{2023arXiv231100432C}  specifically predict that for black holes with masses around $10^7 M_\odot$ accreting at moderate Eddington rates, the MAD state formation and dissolution should occur on yearly timescales, with the exact periodicity modulated by the complex interplay between disk-jet alignment \citep{2020MNRAS.494.3656L}, accretion state transitions \citep{2019A&A...626A.115M} and magnetic field configuration dynamics.  

The two discrete jet ejection episodes captured in 2015--2016 and 2021--2024 offer direct observational support for this model, marking potential recurrence of magnetic energy buildup and release. The relatively short accumulation timescales in Mrk~110 are likely facilitated by its smaller black hole mass and high Eddington ratio, making NLS1 galaxies particularly valuable laboratories for studying transient jet activity and testing magnetically driven jet formation theories. Longer monitoring campaigns are needed to determine any periodicity or recurrence.

These findings provide evidence for dynamic variations within the established classical radio-loud/radio-quiet dichotomy. The discovery of ``changing-look" AGNs transitioning between radio states \citep{2018MNRAS.480.3898N, 2020ApJ...905...74N, 2021MNRAS.502L..61Y, 2025ApJ...979L...2M} supports a continuous distribution in jet properties, likely modulated by episodic conditions rather than fixed intrinsic characteristics. Mrk~110 thus exemplifies a broader class of AGN whose jet activity is intermittent and often escapes detection due to its transient nature and observational constraints.

\subsection{Implications for AGN Classification and Evolution} 

Mrk 110 serves as a transitional object between traditional radio-quiet/loud AGN populations. Despite its (time-averaged) low radio loudness ($R \approx 1.49$), it exhibits superluminal motion comparable to radio-intermediate quasars like III Zw 2 \citep{1996ApJ...471..106F}, which shows similar jet kinematics despite higher radio loudness ($R \approx 200$) \citep{1999ApJ...514L..17F,2000A&A...357L..45B,2023ApJ...944..187W}.

Our observations suggest that the conventional radio-quiet classification captures a temporally averaged state, with some systems capable of intermittent relativistic jet formation.  Within the NLS1 population, Mrk 110 bridges radio-quiet NLS1s and highly beamed $\gamma$-ray emitting NLS1s \citep{2014ApJ...781...75W, 2016AJ....152...12L, 2011A&A...528L..11G}. Its viewing angle is roughly consistent with the nearly pole-on orientations of $\gamma$-ray emitting NLS1s \citep[e.g.][]{2016ApJ...820...52P}, supporting the jet state transition model \citep{2015A&A...575A..13F} where AGN radio properties evolve significantly over time. Analysis across the radio-loudness spectrum reveals a continuous distribution of jet properties (velocities, luminosities, and accretion rates) shaped by both intrinsic (e.g., black hole mass and spin, magnetic field configuration) and geometric factors (viewing angle, Doppler boosting), suggesting that \textit{the apparent bimodality emerges primarily from observational selection effects and the transient nature of jet activity. }

The multi-scale radio structure of Mrk 110 reveals a compelling evolutionary narrative spanning different timescales. The extended kpc-scale emission detected by LOFAR represents the historical jet activity over $10^5$ year timescales, while our VLBI observations capture the currently active relativistic jet confined within a few parsec. This structural hierarchy demonstrates that Mrk 110's jet activity operates through recurrent launching episodes rather than continuous outflow, challenging conventional understanding of radio-quiet AGN feedback mechanisms.

The confinement of the current jet at the transition zone between the BLR ($10^{16} - 10^{18}$ cm, \citealt{2024A&A...681A.101J}) and the NLR/torus region provides direct evidence for a previously underexplored mode of AGN feedback through ``frustrated'' jets. The calculated jet power  ($P_{\rm jet} \approx 7.4 \times 10^{42}$~erg~s$^{-1}$, derived using the relation from \cite{2007MNRAS.381..589M}) is remarkably similar to the radiative luminosity of the broad-line region \citep[$L_{\rm BLR} \approx 1.1 \times 10^{42}$~erg~s$^{-1}$, ][]{2019ApJ...886...42D}. 
This energy equivalence facilitates efficient coupling through jet-induced radiation pressure, cloud disruption and acceleration, turbulence, and gas entrainment with similar interactions observed in other AGN systems \citep{1994ApJ...432...62A, 1996ApJ...469..554C, 2002ApJ...568..627C}.
Rather than extending to kiloparsec scales, these confined jets significantly influence the nuclear environment through localized energy deposition, potentially regulating accretion and outflow processes on BLR scales \citep[e.g.,][]{2012ApJ...760...77A,2024A&A...685A.122U}. 
The observed deceleration quantifies jet-medium interactions at the BLR/NLR transition zone, with implications for feedback models in radio-quiet systems \citep{2021MNRAS.504.3823W, 2023ApJ...944..187W}.

The episodic nature of the jet activity offers critical insight into the duty cycle of AGN jets, with implications for feedback efficiency calculations in galaxy evolution models. Our detection refines existing paradigms that view AGN radio emission as either persistent (radio-loud) or absent (radio-quiet), instead revealing intermittent jet formation as a fundamental process across the AGN population that may often escape detection due to limited sensitivity or monitoring cadence. Mrk 110 thus serves as a prototype for a potentially numerous population of AGN with intermittent relativistic jet capability that impacts their nuclear environments despite their conventional radio-quiet classification.

\section{Conclusions}

Our comprehensive VLBI monitoring campaign of Mrk 110 from 2015 to 2024 provides definitive evidence of relativistic jet formation in this radio-quiet NLS1 galaxy. Key findings include:

\begin{enumerate}
    \item \textbf{Superluminal motion:} Direct measurement of superluminal motion with $\beta_{\rm app} = 1.5$--$3.6$, representing one of the highest velocity outflows detected in a radio-quiet AGN. This motion, combined with spectral evolution from steep ($\alpha \approx -0.63$) to inverted ($\alpha = +0.69$), establishes relativistic jet formation in an AGN traditionally classified as radio quiet.
    
    \item \textbf{Jet deceleration:} Observation of significant jet deceleration at 3--6 pc from the core, coincident with the BLR/NLR transition zone, provides direct evidence for jet--ISM interaction and feedback processes at work in radio-quiet systems.
    
    \item \textbf{Episodic jet launching:} Detection of multiple discrete jet ejection episodes (2015--2016 and 2022--2024) indicates intermittent jet launching that supports dynamic variations within the established classical radio-loud/radio-quiet dichotomy. This episodic nature suggests jet formation occurs when transient conditions align, even in nominally radio-quiet sources.
    
    \item \textbf{Persistent jet orientation:} Co-existence of pc-scale active jets with kpc-scale relics demonstrates persistent jet orientation across vastly different timescales, revealing a consistent central engine mechanism despite intermittent activity.
\end{enumerate}

Our results provide new insights into the classical radio-loud/quiet framework, highlighting that relativistic jets can emerge episodically even in sources classified as radio-quiet.
Mrk 110 exemplifies how traditionally radio-quiet AGN can harbor the fundamental physics required for relativistic outflows when conditions temporarily align. This study connects previously separate AGN categories into a unified evolutionary framework where all supermassive black holes potentially cycle through jet-producing phases of varying duration and intensity, 
with observable radio properties governed by the interplay between intrinsic physical drivers—such as black hole spin, accretion rate, and magnetic flux—and environmental conditions, including gas density, interstellar pressure, and large-scale feedback processes.

Beyond traditional classifications, this work carries profound implications for understanding AGN feedback across cosmic time. The episodic, confined jets we detect likely represent an important energy transfer mechanism operating within nominally radio-quiet systems that dominate the AGN census, complementing our understanding of how AGN influence their host galaxies. This ``hidden" kinetic feedback channel may significantly impact AGN-host galaxy co-evolution models and necessitates a fundamental reassessment of how jets contribute to the energy budget of galactic nuclei, regardless of their classical radio designation. Future theoretical models must incorporate intermittent jet formation as a standard component of AGN evolutionary cycles.

\section{acknowledgments}
This work is supported by the National SKA Program of China (grant number 2022SKA0120102). 
TA acknowledge support from the Xinjiang Tianchi Talent Program, the FAST Special Program (NSFC 12041301). 
SG is supported by the Youth Innovation Promotion Association CAS Program under No. 2021258. 
The VLBI data processing made use of the compute resource of the China SKA Regional Centre \citep{2019NatAs...3.1030A,2022SCPMA..6529501A}. 
The National Radio Astronomy Observatory is a facility of the National Science Foundation operated under cooperative agreement by Associated Universities, Inc. Scientific results from data presented in this publication are derived from the following VLBA project codes: BA114, BW138, BA163 and EVN project code: EW034.
This paper is based in part on data obtained with the LOFAR telescope (LOFAR-ERIC) under project code LC6\_015. LOFAR \citep{2013A&A...556A...2V} is the Low Frequency Array designed and constructed by ASTRON.

\vspace{5mm}
\facilities{VLBA, EVN, LOFAR}
\software{astropy \citep{2013A&A...558A..33A,2018AJ....156..123A}, AIPS \citep{2003ASSL..285..109G}, Difmap \citep{1994BAAS...26..987S} }


\bibliography{sample631}{}

\begin{thebibliography}{}
\expandafter\ifx\csname natexlab\endcsname\relax\def\natexlab#1{#1}\fi
\providecommand{\url}[1]{\href{#1}{#1}}
\providecommand{\dodoi}[1]{doi:~\href{http://doi.org/#1}{\nolinkurl{#1}}}
\providecommand{\doeprint}[1]{\href{http://ascl.net/#1}{\nolinkurl{http://ascl.net/#1}}}
\providecommand{\doarXiv}[1]{\href{https://arxiv.org/abs/#1}{\nolinkurl{https://arxiv.org/abs/#1}}}

\bibitem[{{An} \& {Baan}(2012)}]{2012ApJ...760...77A}
{An}, T., \& {Baan}, W.~A. 2012, \apj, 760, 77,
  \dodoi{10.1088/0004-637X/760/1/77}

\bibitem[{{An} {et~al.}(2022){An}, {Wu}, {Lao}, {Guo}, {Xu}, {Lv}, {Zhang}, \&
  {Zhang}}]{2022SCPMA..6529501A}
{An}, T., {Wu}, X., {Lao}, B., {et~al.} 2022, Science China Physics, Mechanics,
  and Astronomy, 65, 129501, \dodoi{10.1007/s11433-022-1981-8}

\bibitem[{{An} {et~al.}(2019){An}, {Wu}, \& {Hong}}]{2019NatAs...3.1030A}
{An}, T., {Wu}, X.-P., \& {Hong}, X. 2019, Nature Astronomy, 3, 1030,
  \dodoi{10.1038/s41550-019-0943-4}

\bibitem[{{Arav} {et~al.}(1994){Arav}, {Li}, \&
  {Begelman}}]{1994ApJ...432...62A}
{Arav}, N., {Li}, Z.-Y., \& {Begelman}, M.~C. 1994, \apj, 432, 62,
  \dodoi{10.1086/174549}

\bibitem[{{Astropy Collaboration} {et~al.}(2013){Astropy Collaboration},
  {Robitaille}, {Tollerud}, {Greenfield}, {Droettboom}, {Bray}, {Aldcroft},
  {Davis}, {Ginsburg}, {Price-Whelan}, {Kerzendorf}, {Conley}, {Crighton},
  {Barbary}, {Muna}, {Ferguson}, {Grollier}, {Parikh}, {Nair}, {Unther},
  {Deil}, {Woillez}, {Conseil}, {Kramer}, {Turner}, {Singer}, {Fox}, {Weaver},
  {Zabalza}, {Edwards}, {Azalee Bostroem}, {Burke}, {Casey}, {Crawford},
  {Dencheva}, {Ely}, {Jenness}, {Labrie}, {Lim}, {Pierfederici}, {Pontzen},
  {Ptak}, {Refsdal}, {Servillat}, \& {Streicher}}]{2013A&A...558A..33A}
{Astropy Collaboration}, {Robitaille}, T.~P., {Tollerud}, E.~J., {et~al.} 2013,
  \aap, 558, A33, \dodoi{10.1051/0004-6361/201322068}

\bibitem[{{Astropy Collaboration} {et~al.}(2018){Astropy Collaboration},
  {Price-Whelan}, {Sip{\H{o}}cz}, {G{\"u}nther}, {Lim}, {Crawford}, {Conseil},
  {Shupe}, {Craig}, {Dencheva}, {Ginsburg}, {VanderPlas}, {Bradley},
  {P{\'e}rez-Su{\'a}rez}, {de Val-Borro}, {Aldcroft}, {Cruz}, {Robitaille},
  {Tollerud}, {Ardelean}, {Babej}, {Bach}, {Bachetti}, {Bakanov}, {Bamford},
  {Barentsen}, {Barmby}, {Baumbach}, {Berry}, {Biscani}, {Boquien}, {Bostroem},
  {Bouma}, {Brammer}, {Bray}, {Breytenbach}, {Buddelmeijer}, {Burke},
  {Calderone}, {Cano Rodr{\'\i}guez}, {Cara}, {Cardoso}, {Cheedella}, {Copin},
  {Corrales}, {Crichton}, {D'Avella}, {Deil}, {Depagne}, {Dietrich}, {Donath},
  {Droettboom}, {Earl}, {Erben}, {Fabbro}, {Ferreira}, {Finethy}, {Fox},
  {Garrison}, {Gibbons}, {Goldstein}, {Gommers}, {Greco}, {Greenfield},
  {Groener}, {Grollier}, {Hagen}, {Hirst}, {Homeier}, {Horton}, {Hosseinzadeh},
  {Hu}, {Hunkeler}, {Ivezi{\'c}}, {Jain}, {Jenness}, {Kanarek}, {Kendrew},
  {Kern}, {Kerzendorf}, {Khvalko}, {King}, {Kirkby}, {Kulkarni}, {Kumar},
  {Lee}, {Lenz}, {Littlefair}, {Ma}, {Macleod}, {Mastropietro}, {McCully},
  {Montagnac}, {Morris}, {Mueller}, {Mumford}, {Muna}, {Murphy}, {Nelson},
  {Nguyen}, {Ninan}, {N{\"o}the}, {Ogaz}, {Oh}, {Parejko}, {Parley}, {Pascual},
  {Patil}, {Patil}, {Plunkett}, {Prochaska}, {Rastogi}, {Reddy Janga},
  {Sabater}, {Sakurikar}, {Seifert}, {Sherbert}, {Sherwood-Taylor}, {Shih},
  {Sick}, {Silbiger}, {Singanamalla}, {Singer}, {Sladen}, {Sooley},
  {Sornarajah}, {Streicher}, {Teuben}, {Thomas}, {Tremblay}, {Turner},
  {Terr{\'o}n}, {van Kerkwijk}, {de la Vega}, {Watkins}, {Weaver}, {Whitmore},
  {Woillez}, {Zabalza}, \& {Astropy Contributors}}]{2018AJ....156..123A}
{Astropy Collaboration}, {Price-Whelan}, A.~M., {Sip{\H{o}}cz}, B.~M., {et~al.}
  2018, \aj, 156, 123, \dodoi{10.3847/1538-3881/aabc4f}

\bibitem[{{Berger} {et~al.}(2012){Berger}, {Zauderer}, {Pooley}, {Soderberg},
  {Sari}, {Brunthaler}, \& {Bietenholz}}]{2012ApJ...748...36B}
{Berger}, E., {Zauderer}, A., {Pooley}, G.~G., {et~al.} 2012, \apj, 748, 36,
  \dodoi{10.1088/0004-637X/748/1/36}

\bibitem[{{Berton} {et~al.}(2020){Berton}, {J{\"a}rvel{\"a}}, {Crepaldi},
  {L{\"a}hteenm{\"a}ki}, {Tornikoski}, {Congiu}, {Kharb}, {Terreran}, \&
  {Vietri}}]{2020A&A...636A..64B}
{Berton}, M., {J{\"a}rvel{\"a}}, E., {Crepaldi}, L., {et~al.} 2020, \aap, 636,
  A64, \dodoi{10.1051/0004-6361/202037793}

\bibitem[{{Boroson} \& {Green}(1992)}]{1992ApJS...80..109B}
{Boroson}, T.~A., \& {Green}, R.~F. 1992, \apjs, 80, 109,
  \dodoi{10.1086/191661}

\bibitem[{{Brunthaler} {et~al.}(2000){Brunthaler}, {Falcke}, {Bower}, {Aller},
  {Aller}, {Ter{\"a}sranta}, {Lobanov}, {Krichbaum}, \&
  {Patnaik}}]{2000A&A...357L..45B}
{Brunthaler}, A., {Falcke}, H., {Bower}, G.~C., {et~al.} 2000, \aap, 357, L45,
  \dodoi{10.48550/arXiv.astro-ph/0004256}

\bibitem[{{Capetti} {et~al.}(1996){Capetti}, {Axon}, {Macchetto}, {Sparks}, \&
  {Boksenberg}}]{1996ApJ...469..554C}
{Capetti}, A., {Axon}, D.~J., {Macchetto}, F., {Sparks}, W.~B., \&
  {Boksenberg}, A. 1996, \apj, 469, 554, \dodoi{10.1086/177804}

\bibitem[{{Cecil} {et~al.}(2002){Cecil}, {Dopita}, {Groves}, {Wilson},
  {Ferruit}, {P{\'e}contal}, \& {Binette}}]{2002ApJ...568..627C}
{Cecil}, G., {Dopita}, M.~A., {Groves}, B., {et~al.} 2002, \apj, 568, 627,
  \dodoi{10.1086/338950}

\bibitem[{{Cendes} {et~al.}(2021){Cendes}, {Eftekhari}, {Berger}, \&
  {Polisensky}}]{2021ApJ...908..125C}
{Cendes}, Y., {Eftekhari}, T., {Berger}, E., \& {Polisensky}, E. 2021, \apj,
  908, 125, \dodoi{10.3847/1538-4357/abd323}

\bibitem[{{Chatterjee} {et~al.}(2023){Chatterjee}, {Liska}, {Tchekhovskoy}, \&
  {Markoff}}]{2023arXiv231100432C}
{Chatterjee}, K., {Liska}, M., {Tchekhovskoy}, A., \& {Markoff}, S. 2023, arXiv
  e-prints, arXiv:2311.00432, \dodoi{10.48550/arXiv.2311.00432}

\bibitem[{{Czerny} {et~al.}(2009){Czerny}, {Siemiginowska}, {Janiuk},
  {Nikiel-Wroczy{\'n}ski}, \& {Stawarz}}]{2009ApJ...698..840C}
{Czerny}, B., {Siemiginowska}, A., {Janiuk}, A., {Nikiel-Wroczy{\'n}ski}, B.,
  \& {Stawarz}, {\L}. 2009, \apj, 698, 840, \dodoi{10.1088/0004-637X/698/1/840}

\bibitem[{{Dasgupta} \& {Rao}(2006)}]{2006ApJ...651L..13D}
{Dasgupta}, S., \& {Rao}, A.~R. 2006, \apjl, 651, L13, \dodoi{10.1086/509117}

\bibitem[{{Deller} {et~al.}(2011){Deller}, {Brisken}, {Phillips}, {Morgan},
  {Alef}, {Cappallo}, {Middelberg}, {Romney}, {Rottmann}, {Tingay}, \&
  {Wayth}}]{2011PASP..123..275D}
{Deller}, A.~T., {Brisken}, W.~F., {Phillips}, C.~J., {et~al.} 2011, \pasp,
  123, 275, \dodoi{10.1086/658907}

\bibitem[{{Doi} {et~al.}(2013){Doi}, {Asada}, {Fujisawa}, {Nagai}, {Hagiwara},
  {Wajima}, \& {Inoue}}]{2013ApJ...765...69D}
{Doi}, A., {Asada}, K., {Fujisawa}, K., {et~al.} 2013, \apj, 765, 69,
  \dodoi{10.1088/0004-637X/765/1/69}

\bibitem[{{Du} \& {Wang}(2019)}]{2019ApJ...886...42D}
{Du}, P., \& {Wang}, J.-M. 2019, \apj, 886, 42,
  \dodoi{10.3847/1538-4357/ab4908}

\bibitem[{{Falcke} {et~al.}(1996){Falcke}, {Sherwood}, \&
  {Patnaik}}]{1996ApJ...471..106F}
{Falcke}, H., {Sherwood}, W., \& {Patnaik}, A.~R. 1996, \apj, 471, 106,
  \dodoi{10.1086/177956}

\bibitem[{{Falcke} {et~al.}(1999){Falcke}, {Bower}, {Lobanov}, {Krichbaum},
  {Patnaik}, {Aller}, {Aller}, {Ter{\"a}sranta}, {Wright}, \&
  {Sandell}}]{1999ApJ...514L..17F}
{Falcke}, H., {Bower}, G.~C., {Lobanov}, A.~P., {et~al.} 1999, \apjl, 514, L17,
  \dodoi{10.1086/311937}

\bibitem[{{Foschini} {et~al.}(2015){Foschini}, {Berton}, {Caccianiga}, {Ciroi},
  {Cracco}, {Peterson}, {Angelakis}, {Braito}, {Fuhrmann}, {Gallo}, {Grupe},
  {J{\"a}rvel{\"a}}, {Kaufmann}, {Komossa}, {Kovalev}, {L{\"a}hteenm{\"a}ki},
  {Lisakov}, {Lister}, {Mathur}, {Richards}, {Romano}, {Sievers},
  {Tagliaferri}, {Tammi}, {Tibolla}, {Tornikoski}, {Vercellone}, {La Mura},
  {Maraschi}, \& {Rafanelli}}]{2015A&A...575A..13F}
{Foschini}, L., {Berton}, M., {Caccianiga}, A., {et~al.} 2015, \aap, 575, A13,
  \dodoi{10.1051/0004-6361/201424972}

\bibitem[{{Gaia Collaboration}(2022)}]{2022yCat.1355....0G}
{Gaia Collaboration}. 2022, {VizieR Online Data Catalog: Gaia DR3 Part 1. Main
  source (Gaia Collaboration, 2022)}, VizieR On-line Data Catalog: I/355.
  Originally published in: Astron. Astrophys., in prep. (2022),
  \dodoi{10.26093/cds/vizier.1355}

\bibitem[{{Ghisellini} {et~al.}(1993){Ghisellini}, {Padovani}, {Celotti}, \&
  {Maraschi}}]{1993ApJ...407...65G}
{Ghisellini}, G., {Padovani}, P., {Celotti}, A., \& {Maraschi}, L. 1993, \apj,
  407, 65, \dodoi{10.1086/172493}

\bibitem[{{Girdhar} {et~al.}(2022){Girdhar}, {Harrison}, {Mainieri}, {Bittner},
  {Costa}, {Kharb}, {Mukherjee}, {Arrigoni Battaia}, {Alexander}, {Calistro
  Rivera}, {Circosta}, {De Breuck}, {Edge}, {Farina}, {Kakkad}, {Lansbury},
  {Molyneux}, {Mullaney}, {Silpa}, {Thomson}, \& {Ward}}]{2022MNRAS.512.1608G}
{Girdhar}, A., {Harrison}, C.~M., {Mainieri}, V., {et~al.} 2022, \mnras, 512,
  1608, \dodoi{10.1093/mnras/stac073}

\bibitem[{{Giroletti} {et~al.}(2011){Giroletti}, {Paragi}, {Bignall}, {Doi},
  {Foschini}, {Gab{\'a}nyi}, {Reynolds}, {Blanchard}, {Campbell}, {Colomer},
  {Hong}, {Kadler}, {Kino}, {van Langevelde}, {Nagai}, {Phillips}, {Sekido},
  {Szomoru}, \& {Tzioumis}}]{2011A&A...528L..11G}
{Giroletti}, M., {Paragi}, Z., {Bignall}, H., {et~al.} 2011, \aap, 528, L11,
  \dodoi{10.1051/0004-6361/201116639}

\bibitem[{{Greisen}(2003)}]{2003ASSL..285..109G}
{Greisen}, E.~W. 2003, in Astrophysics and Space Science Library, Vol. 285,
  Information Handling in Astronomy - Historical Vistas, ed. A.~{Heck}, 109,
  \dodoi{10.1007/0-306-48080-8_7}

\bibitem[{{Hardcastle} {et~al.}(2007){Hardcastle}, {Evans}, \&
  {Croston}}]{2007MNRAS.376.1849H}
{Hardcastle}, M.~J., {Evans}, D.~A., \& {Croston}, J.~H. 2007, \mnras, 376,
  1849, \dodoi{10.1111/j.1365-2966.2007.11572.x}

\bibitem[{{Jacquemin-Ide} {et~al.}(2024){Jacquemin-Ide}, {Rincon},
  {Tchekhovskoy}, \& {Liska}}]{2024MNRAS.532.1522J}
{Jacquemin-Ide}, J., {Rincon}, F., {Tchekhovskoy}, A., \& {Liska}, M. 2024,
  \mnras, 532, 1522, \dodoi{10.1093/mnras/stae1538}

\bibitem[{{J{\"a}rvel{\"a}} {et~al.}(2022){J{\"a}rvel{\"a}}, {Dahale},
  {Crepaldi}, {Berton}, {Congiu}, \& {Antonucci}}]{2022A&A...658A..12J}
{J{\"a}rvel{\"a}}, E., {Dahale}, R., {Crepaldi}, L., {et~al.} 2022, \aap, 658,
  A12, \dodoi{10.1051/0004-6361/202141698}

\bibitem[{{Jur{\'a}{\v{n}}ov{\'a}} {et~al.}(2024){Jur{\'a}{\v{n}}ov{\'a}},
  {Costantini}, {Di Gesu}, {Ebrero}, {Kaastra}, {Korista}, {Kriss},
  {Mehdipour}, {Piconcelli}, \& {Rogantini}}]{2024A&A...681A.101J}
{Jur{\'a}{\v{n}}ov{\'a}}, A., {Costantini}, E., {Di Gesu}, L., {et~al.} 2024,
  \aap, 681, A101, \dodoi{10.1051/0004-6361/202348067}

\bibitem[{{Kaaz} {et~al.}(2023){Kaaz}, {Murguia-Berthier}, {Chatterjee},
  {Liska}, \& {Tchekhovskoy}}]{2023ApJ...950...31K}
{Kaaz}, N., {Murguia-Berthier}, A., {Chatterjee}, K., {Liska}, M. T.~P., \&
  {Tchekhovskoy}, A. 2023, \apj, 950, 31, \dodoi{10.3847/1538-4357/acc7a1}

\bibitem[{{Kellermann} {et~al.}(1989){Kellermann}, {Sramek}, {Schmidt},
  {Shaffer}, \& {Green}}]{1989AJ.....98.1195K}
{Kellermann}, K.~I., {Sramek}, R., {Schmidt}, M., {Shaffer}, D.~B., \& {Green},
  R. 1989, \aj, 98, 1195, \dodoi{10.1086/115207}

\bibitem[{{Kellermann} {et~al.}(1994){Kellermann}, {Sramek}, {Schmidt},
  {Green}, \& {Shaffer}}]{1994AJ....108.1163K}
{Kellermann}, K.~I., {Sramek}, R.~A., {Schmidt}, M., {Green}, R.~F., \&
  {Shaffer}, D.~B. 1994, \aj, 108, 1163, \dodoi{10.1086/117145}

\bibitem[{{Kellermann} \& {Verschuur}(1988)}]{1988gera.book.....K}
{Kellermann}, K.~I., \& {Verschuur}, G.~L. 1988, {Galactic and Extragalactic
  Radio Astronomy}

\bibitem[{{Kollatschny}(2003)}]{2003A&A...407..461K}
{Kollatschny}, W. 2003, \aap, 407, 461, \dodoi{10.1051/0004-6361:20030928}

\bibitem[{{Kukula} {et~al.}(1998){Kukula}, {Dunlop}, {Hughes}, \&
  {Rawlings}}]{1998MNRAS.297..366K}
{Kukula}, M.~J., {Dunlop}, J.~S., {Hughes}, D.~H., \& {Rawlings}, S. 1998,
  \mnras, 297, 366, \dodoi{10.1046/j.1365-8711.1998.01481.x}

\bibitem[{{Liska} {et~al.}(2020){Liska}, {Tchekhovskoy}, \&
  {Quataert}}]{2020MNRAS.494.3656L}
{Liska}, M., {Tchekhovskoy}, A., \& {Quataert}, E. 2020, \mnras, 494, 3656,
  \dodoi{10.1093/mnras/staa955}

\bibitem[{{Lister} {et~al.}(2009){Lister}, {Cohen}, {Homan}, {Kadler},
  {Kellermann}, {Kovalev}, {Ros}, {Savolainen}, \&
  {Zensus}}]{2009AJ....138.1874L}
{Lister}, M.~L., {Cohen}, M.~H., {Homan}, D.~C., {et~al.} 2009, \aj, 138, 1874,
  \dodoi{10.1088/0004-6256/138/6/1874}

\bibitem[{{Lister} {et~al.}(2016){Lister}, {Aller}, {Aller}, {Homan},
  {Kellermann}, {Kovalev}, {Pushkarev}, {Richards}, {Ros}, \&
  {Savolainen}}]{2016AJ....152...12L}
{Lister}, M.~L., {Aller}, M.~F., {Aller}, H.~D., {et~al.} 2016, \aj, 152, 12,
  \dodoi{10.3847/0004-6256/152/1/12}

\bibitem[{{Marcel} {et~al.}(2019){Marcel}, {Ferreira}, {Clavel}, {Petrucci},
  {Malzac}, {Corbel}, {Rodriguez}, {Belmont}, {Coriat}, {Henri}, \&
  {Cangemi}}]{2019A&A...626A.115M}
{Marcel}, G., {Ferreira}, J., {Clavel}, M., {et~al.} 2019, \aap, 626, A115,
  \dodoi{10.1051/0004-6361/201935060}

\bibitem[{{Marscher} \& {Gear}(1985)}]{1985ApJ...298..114M}
{Marscher}, A.~P., \& {Gear}, W.~K. 1985, \apj, 298, 114,
  \dodoi{10.1086/163592}

\bibitem[{{Mart{\'\i}-Vidal} {et~al.}(2012){Mart{\'\i}-Vidal},
  {P{\'e}rez-Torres}, \& {Lobanov}}]{2012A&A...541A.135M}
{Mart{\'\i}-Vidal}, I., {P{\'e}rez-Torres}, M.~A., \& {Lobanov}, A.~P. 2012,
  \aap, 541, A135, \dodoi{10.1051/0004-6361/201118334}

\bibitem[{{Mattila} {et~al.}(2018){Mattila}, {P{\'e}rez-Torres}, {Efstathiou},
  {Mimica}, {Fraser}, {Kankare}, {Alberdi}, {Aloy}, {Heikkil{\"a}}, {Jonker},
  {Lundqvist}, {Mart{\'\i}-Vidal}, {Meikle}, {Romero-Ca{\~n}izales}, {Smartt},
  {Tsygankov}, {Varenius}, {Alonso-Herrero}, {Bondi}, {Fransson},
  {Herrero-Illana}, {Kangas}, {Kotak}, {Ram{\'\i}rez-Olivencia},
  {V{\"a}is{\"a}nen}, {Beswick}, {Clements}, {Greimel}, {Harmanen},
  {Kotilainen}, {Nandra}, {Reynolds}, {Ryder}, {Walton}, {Wiik}, \&
  {{\"O}stlin}}]{2018Sci...361..482M}
{Mattila}, S., {P{\'e}rez-Torres}, M., {Efstathiou}, A., {et~al.} 2018,
  Science, 361, 482, \dodoi{10.1126/science.aao4669}

\bibitem[{{McKinney} {et~al.}(2012){McKinney}, {Tchekhovskoy}, \&
  {Blandford}}]{2012MNRAS.423.3083M}
{McKinney}, J.~C., {Tchekhovskoy}, A., \& {Blandford}, R.~D. 2012, \mnras, 423,
  3083, \dodoi{10.1111/j.1365-2966.2012.21074.x}

\bibitem[{{Merloni} \& {Heinz}(2007)}]{2007MNRAS.381..589M}
{Merloni}, A., \& {Heinz}, S. 2007, \mnras, 381, 589,
  \dodoi{10.1111/j.1365-2966.2007.12253.x}

\bibitem[{{Meyer} {et~al.}(2024){Meyer}, {Laha}, {Shuvo}, {Roychowdhury},
  {Green}, {Rhodes}, {Hankla}, {Philippov}, {Mbarek}, {laor}, {Begelman},
  {Sadaula}, {Ghosh}, {Bruni}, {Panessa}, {Guainazzi}, {Behar}, {Masterson},
  {Zhang}, {Yang}, {Gurwell}, {Keating}, {Williams-Baldwin}, {Bray},
  {Bempong-Manful}, {Wrigley}, {Bianchi}, {Ricci}, {La Franca}, {Kara},
  {Georganopoulos}, {Oates}, {Nicholl}, {Pal}, \&
  {Cenko}}]{2024arXiv240618061M}
{Meyer}, E.~T., {Laha}, S., {Shuvo}, O.~I., {et~al.} 2024, arXiv e-prints,
  arXiv:2406.18061, \dodoi{10.48550/arXiv.2406.18061}

\bibitem[{{Meyer} {et~al.}(2025){Meyer}, {Laha}, {Shuvo}, {Roychowdhury},
  {Green}, {Rhodes}, {Hankla}, {Philippov}, {Mbarek}, {laor}, {Begelman},
  {Sadaula}, {Ghosh}, {Bruni}, {Panessa}, {Guainazzi}, {Behar}, {Masterson},
  {Zhang}, {Yang}, {Gurwell}, {Keating}, {Williams-Baldwin}, {Bray},
  {Bempong-Manful}, {Wrigley}, {Bianchi}, {Ricci}, {La Franca}, {Kara},
  {Georganopoulos}, {Oates}, {Nicholl}, {Pal}, \&
  {Cenko}}]{2025ApJ...979L...2M}
---. 2025, \apjl, 979, L2, \dodoi{10.3847/2041-8213/ad8651}

\bibitem[{{Mohan} {et~al.}(2022){Mohan}, {An}, {Zhang}, {Yang}, {Yang}, \&
  {Wang}}]{2022ApJ...927...74M}
{Mohan}, P., {An}, T., {Zhang}, Y., {et~al.} 2022, \apj, 927, 74,
  \dodoi{10.3847/1538-4357/ac4cb2}

\bibitem[{{Morabito} {et~al.}(2022){Morabito}, {Jackson}, {Mooney}, {Sweijen},
  {Badole}, {Kukreti}, {Venkattu}, {Groeneveld}, {Kappes}, {Bonnassieux},
  {Drabent}, {Iacobelli}, {Croston}, {Best}, {Bondi}, {Callingham}, {Conway},
  {Deller}, {Hardcastle}, {McKean}, {Miley}, {Moldon}, {R{\"o}ttgering},
  {Tasse}, {Shimwell}, {van Weeren}, {Anderson}, {Asgekar}, {Avruch}, {van
  Bemmel}, {Bentum}, {Bonafede}, {Brouw}, {Butcher}, {Ciardi}, {Corstanje},
  {Coolen}, {Damstra}, {de Gasperin}, {Duscha}, {Eisl{\"o}ffel}, {Engels},
  {Falcke}, {Garrett}, {Griessmeier}, {Gunst}, {van Haarlem}, {Hoeft}, {van der
  Horst}, {J{\"u}tte}, {Kadler}, {Koopmans}, {Krankowski}, {Mann}, {Nelles},
  {Oonk}, {Orru}, {Paas}, {Pandey}, {Pizzo}, {Pandey-Pommier}, {Reich},
  {Rothkaehl}, {Ruiter}, {Schwarz}, {Shulevski}, {Soida}, {Tagger}, {Vocks},
  {Wijers}, {Wijnholds}, {Wucknitz}, {Zarka}, \& {Zucca}}]{2022A&A...658A...1M}
{Morabito}, L.~K., {Jackson}, N.~J., {Mooney}, S., {et~al.} 2022, \aap, 658,
  A1, \dodoi{10.1051/0004-6361/202140649}

\bibitem[{{Mukherjee} {et~al.}(2016){Mukherjee}, {Bicknell}, {Sutherland}, \&
  {Wagner}}]{2016MNRAS.461..967M}
{Mukherjee}, D., {Bicknell}, G.~V., {Sutherland}, R., \& {Wagner}, A. 2016,
  \mnras, 461, 967, \dodoi{10.1093/mnras/stw1368}

\bibitem[{{Mukherjee} {et~al.}(2018){Mukherjee}, {Bicknell}, {Wagner},
  {Sutherland}, \& {Silk}}]{2018MNRAS.479.5544M}
{Mukherjee}, D., {Bicknell}, G.~V., {Wagner}, A.~Y., {Sutherland}, R.~S., \&
  {Silk}, J. 2018, \mnras, 479, 5544, \dodoi{10.1093/mnras/sty1776}

\bibitem[{{Netzer}(2015)}]{2015ARA&A..53..365N}
{Netzer}, H. 2015, \araa, 53, 365, \dodoi{10.1146/annurev-astro-082214-122302}

\bibitem[{{Noda} \& {Done}(2018)}]{2018MNRAS.480.3898N}
{Noda}, H., \& {Done}, C. 2018, \mnras, 480, 3898,
  \dodoi{10.1093/mnras/sty2032}

\bibitem[{{Nyland} {et~al.}(2020){Nyland}, {Dong}, {Patil}, {Lacy}, {van
  Velzen}, {Kimball}, {Sarbadhicary}, {Hallinan}, {Baldassare}, {Clarke},
  {Goulding}, {Greene}, {Hughes}, {Kassim}, {Kunert-Bajraszewska}, {Maccarone},
  {Mooley}, {Mukherjee}, {Peters}, {Petrov}, {Polisensky}, {Rujopakarn},
  {Whittle}, \& {Vaccari}}]{2020ApJ...905...74N}
{Nyland}, K., {Dong}, D.~Z., {Patil}, P., {et~al.} 2020, \apj, 905, 74,
  \dodoi{10.3847/1538-4357/abc341}

\bibitem[{{Osterbrock} \& {Pogge}(1985)}]{1985ApJ...297..166O}
{Osterbrock}, D.~E., \& {Pogge}, R.~W. 1985, \apj, 297, 166,
  \dodoi{10.1086/163513}

\bibitem[{{Paliya} \& {Stalin}(2016)}]{2016ApJ...820...52P}
{Paliya}, V.~S., \& {Stalin}, C.~S. 2016, \apj, 820, 52,
  \dodoi{10.3847/0004-637X/820/1/52}

\bibitem[{{Panessa} {et~al.}(2022){Panessa}, {P{\'e}rez-Torres},
  {Hern{\'a}ndez-Garc{\'\i}a}, {Casella}, {Giroletti}, {Orienti}, {Baldi},
  {Bassani}, {Fiocchi}, {La Franca}, {Malizia}, {McHardy}, {Nicastro}, {Piro},
  {Vincentelli}, {Williams}, \& {Ubertini}}]{2022MNRAS.510..718P}
{Panessa}, F., {P{\'e}rez-Torres}, M., {Hern{\'a}ndez-Garc{\'\i}a}, L.,
  {et~al.} 2022, \mnras, 510, 718, \dodoi{10.1093/mnras/stab3426}

\bibitem[{{Peterson} {et~al.}(2004){Peterson}, {Ferrarese}, {Gilbert}, {Kaspi},
  {Malkan}, {Maoz}, {Merritt}, {Netzer}, {Onken}, {Pogge}, {Vestergaard}, \&
  {Wandel}}]{2004ApJ...613..682P}
{Peterson}, B.~M., {Ferrarese}, L., {Gilbert}, K.~M., {et~al.} 2004, \apj, 613,
  682, \dodoi{10.1086/423269}

\bibitem[{{Porquet} {et~al.}(2024){Porquet}, {Hagen}, {Grosso}, {Lobban},
  {Reeves}, {Braito}, \& {Done}}]{2024A&A...681A..40P}
{Porquet}, D., {Hagen}, S., {Grosso}, N., {et~al.} 2024, \aap, 681, A40,
  \dodoi{10.1051/0004-6361/202347202}

\bibitem[{{Porquet} {et~al.}(2021){Porquet}, {Reeves}, {Grosso}, {Braito}, \&
  {Lobban}}]{2021A&A...654A..89P}
{Porquet}, D., {Reeves}, J.~N., {Grosso}, N., {Braito}, V., \& {Lobban}, A.
  2021, \aap, 654, A89, \dodoi{10.1051/0004-6361/202141577}

\bibitem[{{Pudritz} {et~al.}(2012){Pudritz}, {Hardcastle}, \&
  {Gabuzda}}]{2012SSRv..169...27P}
{Pudritz}, R.~E., {Hardcastle}, M.~J., \& {Gabuzda}, D.~C. 2012, \ssr, 169, 27,
  \dodoi{10.1007/s11214-012-9895-z}

\bibitem[{{Salvesen} {et~al.}(2016){Salvesen}, {Simon}, {Armitage}, \&
  {Begelman}}]{2016MNRAS.457..857S}
{Salvesen}, G., {Simon}, J.~B., {Armitage}, P.~J., \& {Begelman}, M.~C. 2016,
  \mnras, 457, 857, \dodoi{10.1093/mnras/stw029}

\bibitem[{{Schinzel} {et~al.}(2012){Schinzel}, {Lobanov}, {Taylor}, {Jorstad},
  {Marscher}, \& {Zensus}}]{2012A&A...537A..70S}
{Schinzel}, F.~K., {Lobanov}, A.~P., {Taylor}, G.~B., {et~al.} 2012, \aap, 537,
  A70, \dodoi{10.1051/0004-6361/201117705}

\bibitem[{{Shepherd} {et~al.}(1994){Shepherd}, {Pearson}, \&
  {Taylor}}]{1994BAAS...26..987S}
{Shepherd}, M.~C., {Pearson}, T.~J., \& {Taylor}, G.~B. 1994, in Bulletin of
  the American Astronomical Society, Vol.~26, 987--989

\bibitem[{{Sikora} \& {Begelman}(2013)}]{2013ApJ...764L..24S}
{Sikora}, M., \& {Begelman}, M.~C. 2013, \apjl, 764, L24,
  \dodoi{10.1088/2041-8205/764/2/L24}

\bibitem[{{Tchekhovskoy} {et~al.}(2011){Tchekhovskoy}, {Narayan}, \&
  {McKinney}}]{2011MNRAS.418L..79T}
{Tchekhovskoy}, A., {Narayan}, R., \& {McKinney}, J.~C. 2011, \mnras, 418, L79,
  \dodoi{10.1111/j.1745-3933.2011.01147.x}

\bibitem[{{Ulivi} {et~al.}(2024){Ulivi}, {Venturi}, {Cresci}, {Marconi},
  {Marconcini}, {Amiri}, {Belfiore}, {Bertola}, {Carniani}, {D'Amato}, {Di
  Teodoro}, {Ginolfi}, {Girdhar}, {Harrison}, {Maiolino}, {Mannucci},
  {Mingozzi}, {Perna}, {Scialpi}, {Tomicic}, {Tozzi}, \&
  {Treister}}]{2024A&A...685A.122U}
{Ulivi}, L., {Venturi}, G., {Cresci}, G., {et~al.} 2024, \aap, 685, A122,
  \dodoi{10.1051/0004-6361/202347436}

\bibitem[{{van Haarlem} {et~al.}(2013){van Haarlem}, {Wise}, {Gunst}, {Heald},
  {McKean}, {Hessels}, {de Bruyn}, {Nijboer}, {Swinbank}, {Fallows},
  {Brentjens}, {Nelles}, {Beck}, {Falcke}, {Fender}, {H{\"o}randel},
  {Koopmans}, {Mann}, {Miley}, {R{\"o}ttgering}, {Stappers}, {Wijers},
  {Zaroubi}, {van den Akker}, {Alexov}, {Anderson}, {Anderson}, {van Ardenne},
  {Arts}, {Asgekar}, {Avruch}, {Batejat}, {B{\"a}hren}, {Bell}, {Bell}, {van
  Bemmel}, {Bennema}, {Bentum}, {Bernardi}, {Best}, {B{\^\i}rzan}, {Bonafede},
  {Boonstra}, {Braun}, {Bregman}, {Breitling}, {van de Brink}, {Broderick},
  {Broekema}, {Brouw}, {Br{\"u}ggen}, {Butcher}, {van Cappellen}, {Ciardi},
  {Coenen}, {Conway}, {Coolen}, {Corstanje}, {Damstra}, {Davies}, {Deller},
  {Dettmar}, {van Diepen}, {Dijkstra}, {Donker}, {Doorduin}, {Dromer}, {Drost},
  {van Duin}, {Eisl{\"o}ffel}, {van Enst}, {Ferrari}, {Frieswijk}, {Gankema},
  {Garrett}, {de Gasperin}, {Gerbers}, {de Geus}, {Grie{\ss}meier}, {Grit},
  {Gruppen}, {Hamaker}, {Hassall}, {Hoeft}, {Holties}, {Horneffer}, {van der
  Horst}, {van Houwelingen}, {Huijgen}, {Iacobelli}, {Intema}, {Jackson},
  {Jelic}, {de Jong}, {Juette}, {Kant}, {Karastergiou}, {Koers}, {Kollen},
  {Kondratiev}, {Kooistra}, {Koopman}, {Koster}, {Kuniyoshi}, {Kramer},
  {Kuper}, {Lambropoulos}, {Law}, {van Leeuwen}, {Lemaitre}, {Loose}, {Maat},
  {Macario}, {Markoff}, {Masters}, {McFadden}, {McKay-Bukowski}, {Meijering},
  {Meulman}, {Mevius}, {Middelberg}, {Millenaar}, {Miller-Jones}, {Mohan},
  {Mol}, {Morawietz}, {Morganti}, {Mulcahy}, {Mulder}, {Munk}, {Nieuwenhuis},
  {van Nieuwpoort}, {Noordam}, {Norden}, {Noutsos}, {Offringa}, {Olofsson},
  {Omar}, {Orr{\'u}}, {Overeem}, {Paas}, {Pandey-Pommier}, {Pandey}, {Pizzo},
  {Polatidis}, {Rafferty}, {Rawlings}, {Reich}, {de Reijer}, {Reitsma},
  {Renting}, {Riemers}, {Rol}, {Romein}, {Roosjen}, {Ruiter}, {Scaife}, {van
  der Schaaf}, {Scheers}, {Schellart}, {Schoenmakers}, {Schoonderbeek},
  {Serylak}, {Shulevski}, {Sluman}, {Smirnov}, {Sobey}, {Spreeuw}, {Steinmetz},
  {Sterks}, {Stiepel}, {Stuurwold}, {Tagger}, {Tang}, {Tasse}, {Thomas},
  {Thoudam}, {Toribio}, {van der Tol}, {Usov}, {van Veelen}, {van der Veen},
  {ter Veen}, {Verbiest}, {Vermeulen}, {Vermaas}, {Vocks}, {Vogt}, {de Vos},
  {van der Wal}, {van Weeren}, {Weggemans}, {Weltevrede}, {White}, {Wijnholds},
  {Wilhelmsson}, {Wucknitz}, {Yatawatta}, {Zarka}, \&
  {Zensus}}]{2013A&A...556A...2V}
{van Haarlem}, M.~P., {Wise}, M.~W., {Gunst}, A.~W., {et~al.} 2013, \aap, 556,
  A2, \dodoi{10.1051/0004-6361/201220873}

\bibitem[{{Vincentelli} {et~al.}(2022){Vincentelli}, {McHardy}, {Hern{\'a}ndez
  Santisteban}, {Cackett}, {Gelbord}, {Horne}, {Miller}, \&
  {Lobban}}]{2022MNRAS.512L..33V}
{Vincentelli}, F.~M., {McHardy}, I., {Hern{\'a}ndez Santisteban}, J.~V.,
  {et~al.} 2022, \mnras, 512, L33, \dodoi{10.1093/mnrasl/slac009}

\bibitem[{{Vincentelli} {et~al.}(2021){Vincentelli}, {McHardy}, {Cackett},
  {Barth}, {Horne}, {Goad}, {Korista}, {Gelbord}, {Brandt}, {Edelson},
  {Miller}, {Pahari}, {Peterson}, {Schmidt}, {Baldi}, {Breedt}, {Hern{\'a}ndez
  Santisteban}, {Romero-Colmenero}, {Ward}, \&
  {Williams}}]{2021MNRAS.504.4337V}
{Vincentelli}, F.~M., {McHardy}, I., {Cackett}, E.~M., {et~al.} 2021, \mnras,
  504, 4337, \dodoi{10.1093/mnras/stab1033}

\bibitem[{{Wajima} {et~al.}(2014){Wajima}, {Fujisawa}, {Hayashida}, {Isobe},
  {Ishida}, \& {Yonekura}}]{2014ApJ...781...75W}
{Wajima}, K., {Fujisawa}, K., {Hayashida}, M., {et~al.} 2014, \apj, 781, 75,
  \dodoi{10.1088/0004-637X/781/2/75}

\bibitem[{{Wang} {et~al.}(2023{\natexlab{a}}){Wang}, {An}, {Cheng}, {Ho},
  {Kellermann}, {Baan}, {Yang}, \& {Zhang}}]{2023MNRAS.518...39W}
{Wang}, A., {An}, T., {Cheng}, X., {et~al.} 2023{\natexlab{a}}, \mnras, 518,
  39, \dodoi{10.1093/mnras/stac3091}

\bibitem[{{Wang} {et~al.}(2021){Wang}, {An}, {Jaiswal}, {Mohan}, {Wang},
  {Baan}, {Zhang}, \& {Yang}}]{2021MNRAS.504.3823W}
{Wang}, A., {An}, T., {Jaiswal}, S., {et~al.} 2021, \mnras, 504, 3823,
  \dodoi{10.1093/mnras/stab587}

\bibitem[{{Wang} {et~al.}(2023{\natexlab{b}}){Wang}, {An}, {Zhang}, {Cheng},
  {Ho}, {Kellermann}, \& {Baan}}]{2023MNRAS.525.6064W}
{Wang}, A., {An}, T., {Zhang}, Y., {et~al.} 2023{\natexlab{b}}, \mnras, 525,
  6064, \dodoi{10.1093/mnras/stad2651}

\bibitem[{{Wang} {et~al.}(2023{\natexlab{c}}){Wang}, {An}, {Guo}, {Ho}, {Baan},
  {Braun}, {Chen}, {Cheng}, {Hartley}, {Yang}, \&
  {Zhang}}]{2023MNRAS.523L..30W}
{Wang}, A., {An}, T., {Guo}, S., {et~al.} 2023{\natexlab{c}}, \mnras, 523, L30,
  \dodoi{10.1093/mnrasl/slad051}

\bibitem[{{Wang} {et~al.}(2023{\natexlab{d}}){Wang}, {An}, {Guo}, {Mohan},
  {Chamani}, {Baan}, {Hovatta}, {Falcke}, {Galvin}, {Hurley-Walker}, {Jaiswal},
  {Lahteenmaki}, {Lao}, {Lv}, {Tornikoski}, \& {Zhang}}]{2023ApJ...944..187W}
---. 2023{\natexlab{d}}, \apj, 944, 187, \dodoi{10.3847/1538-4357/acaf02}

\bibitem[{{Yang} {et~al.}(2021){Yang}, {van Bemmel}, {Paragi}, {Komossa},
  {Yuan}, {Yang}, {An}, {Koay}, {Reynolds}, {Oonk}, {Liu}, \&
  {Wu}}]{2021MNRAS.502L..61Y}
{Yang}, J., {van Bemmel}, I., {Paragi}, Z., {et~al.} 2021, \mnras, 502, L61,
  \dodoi{10.1093/mnrasl/slab005}

\bibitem[{{Zauderer} {et~al.}(2011){Zauderer}, {Berger}, {Soderberg}, {Loeb},
  {Narayan}, {Frail}, {Petitpas}, {Brunthaler}, {Chornock}, {Carpenter},
  {Pooley}, {Mooley}, {Kulkarni}, {Margutti}, {Fox}, {Nakar}, {Patel},
  {Volgenau}, {Culverhouse}, {Bietenholz}, {Rupen}, {Max-Moerbeck}, {Readhead},
  {Richards}, {Shepherd}, {Storm}, \& {Hull}}]{2011Natur.476..425Z}
{Zauderer}, B.~A., {Berger}, E., {Soderberg}, A.~M., {et~al.} 2011, \nat, 476,
  425, \dodoi{10.1038/nature10366}

\end{thebibliography}
\bibliographystyle{aasjournal}

\appendix

\section{VLBI Observations and data processing} \label{app:VLBI}

We conducted an extensive multi-frequency VLBI monitoring campaign of Mrk 110 spanning from 2021 to 2024, employing both the Very Long Baseline Array (VLBA) and the European VLBI Network (EVN). Our observational strategy was designed based on previous successful monitoring campaigns of radio-quiet AGN \citep[e.g.,][]{2023MNRAS.518...39W,2023MNRAS.523L..30W} and optimized to detect compact radio emission at milliarcsecond scales. The observations were primarily conducted at two frequencies (4.7 and 7.6 GHz), chosen to provide both good sensitivity and sufficient spectral coverage to constrain the emission mechanism through in-band spectral index measurements.

Our monitoring campaign began with an exploratory observation in 2015 August using the VLBA at 5.0 GHz, which established the feasibility of detecting and imaging the compact radio emission from Mrk 110. Following the detection of significant variability in the X-ray band \citep{2021MNRAS.504.4337V}, we initiated a more comprehensive monitoring program starting with triple-frequency VLBA observations in 2021 December. The success of these observations, combined with the detection of enhanced activity, motivated an intensive monitoring campaign from 2023 March through 2024 February, incorporating both VLBA and EVN observations to maximize the temporal coverage and sensitivity.

Each VLBA observation lasted approximately 3 hours, with frequent scans on the bright calibrator J0932+5306 for phase calibration. The data were recorded with a 4 Gbps sampling rate in dual polarization, using 4 intermediate frequency (IF) bands of 512 MHz bandwidth each. For the EVN observation in March 2023, we conducted a 6-hour observation at a data rate of 2048 Mbps, involving 9 stations and providing different projected baselines compared to the VLBA observations.
Weather conditions were generally favorable throughout the campaign, with typical system temperatures ranging from 10--38 K for most stations.

\begin{table*}
\centering
\caption{VLBI image parameters. Columns list: (1) observation date, (2) code, (3) frequency (GHz), (4) beam size and position angle, (5) peak flux density and image rms (root-mean-square) noise, (6) integrated flux density, (7) component FWHM (full width at half maximum) size, and (8) brightness temperature. Note: The component FWHM measured on 2023-03-28 is approximately $0.01$ measured by both MCMC and \texttt{DIFMAP}, which is unreasonably small. Therefore, we calculated a more reliable value based on the method described in \citep{2012A&A...537A..70S} and presented it in the table.}
\begin{tabular}{cccclccccc}
\hline \hline 
Obs. date  & Project & Freq.   & $B_{\rm maj}\times B_{\rm min}$, $B_{\rm PA}$  &$S_{\rm peak}$ / $\sigma_{\rm rms}$ &$S_{\rm int}$  &$\theta_{\rm FWHM}$     &log($T_{\rm B}$) \\
(yyyy-mm-dd) &  code   &(GHz)         & (mas$\times$mas, \degr )           &(mJy beam$^{-1}$)   & (mJy)         & (mas)                 & (log K)        \\
(1)       & (2)           & (3)    & (4)       & (5)                                & (6)             & (7)                & (8)                              \\
\hline 
2015-08-04&BA114          &5.0  &4.72$\times$2.01,8.71     &0.83 / 0.027   &1.05$\pm$0.03     &1.03$\pm$0.03    &7.68$\pm$0.24 \\
2015-08-14&BP193          &4.4  &5.05$\times$1.78,$-$1.25  &0.64 / 0.033   &0.81$\pm$0.04     &1.29$\pm$0.07    &7.49$\pm$0.37 \\
2015-08-14&BP193          &7.6  &2.88$\times$1.01,$-$1.79  &0.33 / 0.040   &0.35$\pm$0.04     &0.25$\pm$0.03    &8.07$\pm$0.95 \\
2016-04-24&BP196          &5.0  &4.94$\times$1.40,$-$14.30 &0.79 / 0.024   &0.92$\pm$0.03     &0.76$\pm$0.02    &7.89$\pm$0.23 \\
2016-04-25&BP196          &5.0  &4.83$\times$1.65,$-$19.20 &0.91 / 0.021   &0.99$\pm$0.02     &0.69$\pm$0.02    &8.01$\pm$0.18 \\
2016-04-27&BP196          &5.0  &4.46$\times$1.57,$-$18.60 &0.91 / 0.026   &0.99$\pm$0.03     &0.62$\pm$0.02    &8.10$\pm$0.22 \\
2016-04-28&BP196          &5.0  &4.27$\times$1.42,$-$12.00 &0.73 / 0.021   &0.89$\pm$0.03     &0.91$\pm$0.03    &7.72$\pm$0.22 \\
2016-04-29&BP196          &5.0  &4.40$\times$1.48,$-$12.30 &0.69 / 0.019   &0.89$\pm$0.02     &1.11$\pm$0.03    &7.55$\pm$0.20 \\
2016-04-30&BP196          &5.0  &4.35$\times$1.48,$-$12.30 &0.77 / 0.019   &0.90$\pm$0.02     &0.81$\pm$0.02    &7.83$\pm$0.19 \\
2016-05-01&BP196          &5.0  &4.32$\times$1.48,$-$12.20 &0.75 / 0.019   &0.87$\pm$0.02     &0.78$\pm$0.02    &7.84$\pm$0.19 \\
2016-05-02&BP196          &5.0  &4.37$\times$1.49,$-$12.00 &0.77 / 0.019   &0.91$\pm$0.02     &0.86$\pm$0.02    &7.78$\pm$0.19 \\
2016-05-03&BP196          &5.0  &5.38$\times$1.46,$-$15.20 &0.74 / 0.023   &0.90$\pm$0.03     &0.94$\pm$0.03    &7.70$\pm$0.23 \\
2016-05-04&BP196          &5.0  &5.27$\times$1.32,$-$14.40 &0.68 / 0.024   &0.79$\pm$0.03     &0.69$\pm$0.02    &7.91$\pm$0.27 \\
2016-05-06&BP196          &5.0  &4.72$\times$0.87,$-$12.80 &0.61 / 0.052   &0.80$\pm$0.07     &0.85$\pm$0.07    &7.73$\pm$0.64 \\
2016-05-07&BP196          &5.0  &5.67$\times$1.36,$-$11.40 &0.66 / 0.024   &0.82$\pm$0.03     &0.94$\pm$0.03    &7.66$\pm$0.27 \\
2016-05-08&BP196          &5.0  &5.19$\times$1.91,$-$15.50 &0.78 / 0.023   &1.09$\pm$0.03     &1.66$\pm$0.05    &7.29$\pm$0.21 \\
2016-05-18&BP203          &5.0  &4.01$\times$1.73,26.80    &0.67 / 0.025   &0.78$\pm$0.03     &0.86$\pm$0.03    &7.71$\pm$0.28 \\
2016-05-19&BP203          &5.0  &4.36$\times$1.54,23.90    &0.72 / 0.020   &0.80$\pm$0.02     &0.73$\pm$0.02    &7.86$\pm$0.21 \\
2021-12-31&BW138          &7.6  &2.59$\times$0.86,67.00    &0.69 / 0.036   &0.88$\pm$0.05     &0.58$\pm$0.03    &7.74$\pm$0.39 \\
2021-12-31&BW138          &4.7  &4.23$\times$1.43,68.00    &0.95 / 0.035   &1.12$\pm$0.04     &0.75$\pm$0.03    &8.04$\pm$0.29 \\
2023-03-28&EW034          &4.9  &2.26$\times$0.80,12.00    &0.78 / 0.029   &0.88$\pm$0.03     &0.49$\pm$0.02    &8.27$\pm$0.30 \\
2023-08-03&BA163          &7.6  &2.50$\times$0.92,3.97     &1.05 / 0.026   &1.29$\pm$0.03     &0.59$\pm$0.01    &7.89$\pm$0.19 \\
2023-08-03&BA163          &4.7  &4.15$\times$1.53,2.49     &1.05 / 0.022   &1.20$\pm$0.03     &0.76$\pm$0.02    &8.06$\pm$0.16 \\
2023-11-17&BA163          &7.6  &2.59$\times$1.48,$-$9.64  &1.52 / 0.028   &1.55$\pm$0.03     &0.22$\pm$0.01    &8.83$\pm$0.16 \\
2023-11-17&BA163          &4.7  &4.04$\times$1.92,$-$2.88  &1.08 / 0.023   &1.19$\pm$0.03     &0.76$\pm$0.02    &8.06$\pm$0.17 \\
2024-02-02&BA163          &7.6  &2.74$\times$1.41,2.89     &1.48 / 0.032   &2.03$\pm$0.04     &1.07$\pm$0.02    &7.57$\pm$0.16 \\
2024-02-02&BA163          &4.7  &4.11$\times$1.34,6.73     &1.18 / 0.024   &1.46$\pm$0.03     &0.89$\pm$0.02    &8.01$\pm$0.16 \\ \hline
\end{tabular} 
\label{tab:obs-info}
\end{table*}

The correlation of the VLBA data was performed at the Array Operations Center in Socorro using the DiFX software correlator \citep{2011PASP..123..275D}, while the EVN data were correlated at the Joint Institute for VLBI ERIC (JIVE) using their own SFXC correlator. We employed an integration time of 2 seconds and 32 frequency channels per IF to minimize time and bandwidth smearing effects while maintaining manageable data volumes. This setup provided sufficient resolution to image the inner few parsecs of Mrk 110 without significant loss of sensitivity due to averaging effects.

The data reduction was carried out using the NRAO Astronomical Image Processing System \citep[AIPS;][]{2003ASSL..285..109G}, following procedures optimized for weak sources \citep{2012A&A...541A.135M}. The data analysis was conducted by using the compute resource of the China SKA Regional Centre \citep{2019NatAs...3.1030A,2022SCPMA..6529501A}. We used the standard VLBA data reduction procedures. We first applied gain calibration for the flux density scale using the AIPS task \texttt{ANTAB}, followed by ionospheric corrections using the AIPS task \texttt{TECOR} with total electron content (TEC) maps from the NASA CDDIS archive. The Earth Orientation Parameters (EOP) were corrected using \texttt{CLCOR}, and instrumental delays were identified and removed by running \texttt{VLBAMPCL}.

After initial calibration, we performed global fringe-fitting using the AIPS task \texttt{FRING} to solve for residual delays, rates, and phases. For the weaker epochs, we employed baseline-based fringe-fitting techniques following \citet{2012A&A...541A.135M} to maximize the detection sensitivity. 
Bandpass calibration was performed using observations of bright calibrators J0834+5534, chosen for their high flux density and minimal structure.
The calibrated data were then iteratively imaged and self-calibrated using the \texttt{DIFMAP} package \citep{1994BAAS...26..987S}, applying phase-only self-calibration initially, followed by combined amplitude and phase self-calibration with progressively shorter solution intervals as the signal-to-noise ratio improved. The coordinates were measured by \texttt{DIFMAP} (Table \ref{tab:coordinate}).

The final images were produced using natural weighting in \texttt{DIFMAP} to optimize the trade-off between resolution and sensitivity, following extensive testing of different weighting schemes. This approach yielded typical restoring beam sizes of $4.72 \times 2.01$ mas at 4.7 GHz, with corresponding beams scaling approximately with frequency at the higher bands. The resulting noise levels in the images were typically 20--35 $\mu$Jy/beam, consistent with theoretical expectations given the array sensitivity and integration time. 

\begin{table*}
\centering
\caption{Coordinate. Columns list: (1) observation date, (2) frequency (GHz), (3) coordinate, (4) coordinate errors in the Right Ascension and Declination directions.}
\begin{tabular}{cccc}
\hline \hline 
Obs. date    & Freq.  & RA, Dec                         & $\Delta \delta$ \\
(yyyy-mm-dd) &(GHz)   &(J2000)                          & (mas)           \\  
(1)          & (2)    & (3)                             & (4)             \\
\hline 
2015-08-04   &5.0     &09:25:12.84779, 52:17:10.3862    &0.07              \\
2015-08-14   &4.4     &09:25:12.84777, 52:17:10.3864    &0.10              \\
2015-08-14   &7.6     &09:25:12.84781, 52:17:10.3866    &0.14              \\
2016-04-24   &5.0     &09:25:12.84773, 52:17:10.3876    &0.07              \\
2016-04-25   &5.0     &09:25:12.84776, 52:17:10.3873    &0.05              \\
2016-04-27   &5.0     &09:25:12.84776, 52:17:10.3872    &0.06              \\
2016-04-28   &5.0     &09:25:12.84774, 52:17:10.3875    &0.06              \\
2016-04-29   &5.0     &09:25:12.84773, 52:17:10.3874    &0.06              \\
2016-04-30   &5.0     &09:25:12.84774, 52:17:10.3874    &0.05              \\
2016-05-01   &5.0     &09:25:12.84776, 52:17:10.3873    &0.05              \\
2016-05-02   &5.0     &09:25:12.84774, 52:17:10.3874    &0.05              \\
2016-05-03   &5.0     &09:25:12.84775, 52:17:10.3873    &0.07              \\
2016-05-04   &5.0     &09:25:12.84774, 52:17:10.3877    &0.08              \\
2016-05-06   &5.0     &09:25:12.84773, 52:17:10.3875    &0.15              \\
2016-05-07   &5.0     &09:25:12.84773, 52:17:10.3873    &0.08              \\
2016-05-08   &5.0     &09:25:12.84771, 52:17:10.3878    &0.07              \\
2016-05-18   &5.0     &09:25:12.84776, 52:17:10.3871    &0.07              \\
2016-05-19   &5.0     &09:25:12.84776, 52:17:10.3871    &0.06              \\
2021-12-31   &7.6     &09:25:12.84782, 52:17:10.3861    &0.06              \\
2021-12-31   &4.7     &09:25:12.84779, 52:17:10.3861    &0.07              \\
2023-03-28   &4.9     &09:25:12.84782, 52:17:10.3877    &0.05              \\
2023-08-03   &7.6     &09:25:12.84778, 52:17:10.3874    &0.04              \\
2023-08-03   &4.7     &09:25:12.84781, 52:17:10.3875    &0.05              \\
2023-11-17   &7.6     &09:25:12.84777, 52:17:10.3875    &0.04              \\
2023-11-17   &4.7     &09:25:12.84775, 52:17:10.3874    &0.05              \\
2024-02-02   &7.6     &09:25:12.84776, 52:17:10.3876    &0.04              \\
2024-02-02   &4.7     &09:25:12.84775, 52:17:10.3876    &0.05              \\\hline
     
\end{tabular} 
\label{tab:coordinate}
\end{table*}

\section{VLBI images} \label{app:vlbiimage}

\begin{figure*}
    \centering
    \includegraphics[width=0.25\linewidth]{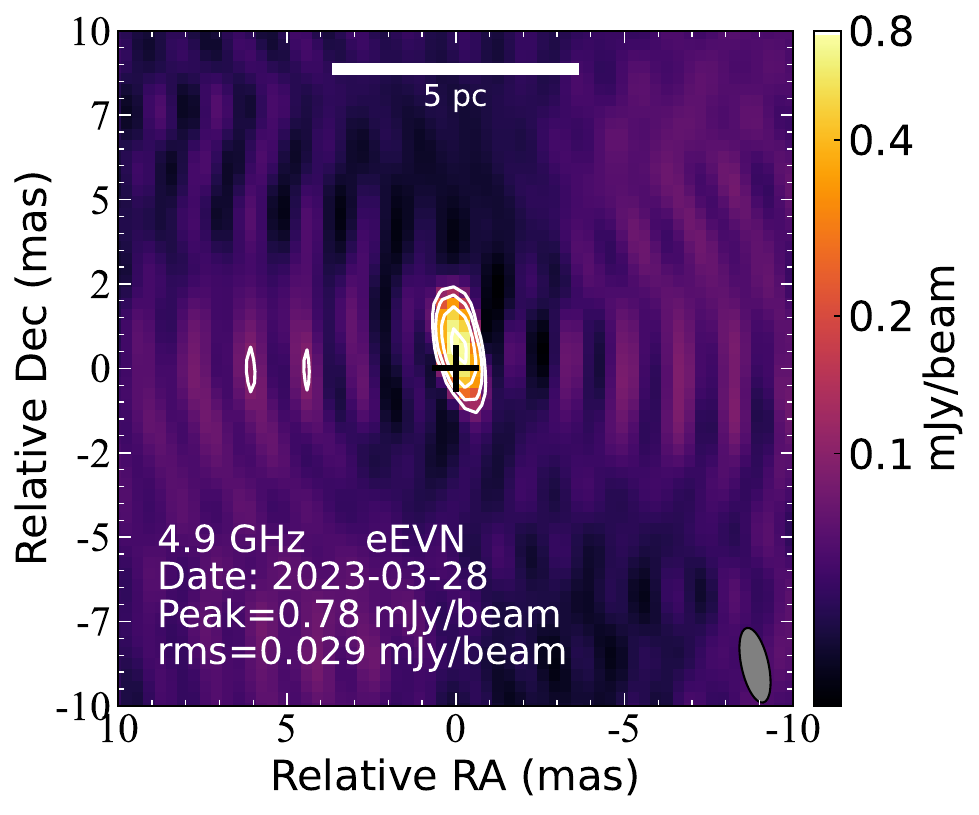}\\
    \includegraphics[width=0.25\linewidth]{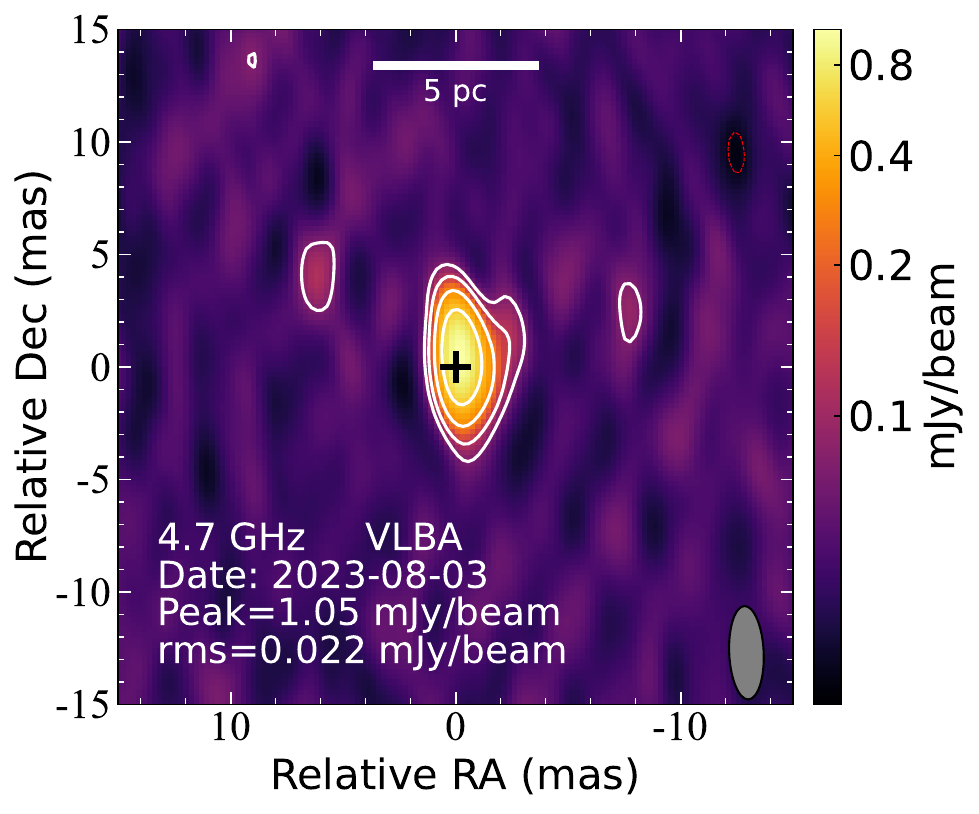}
    \includegraphics[width=0.25\linewidth]{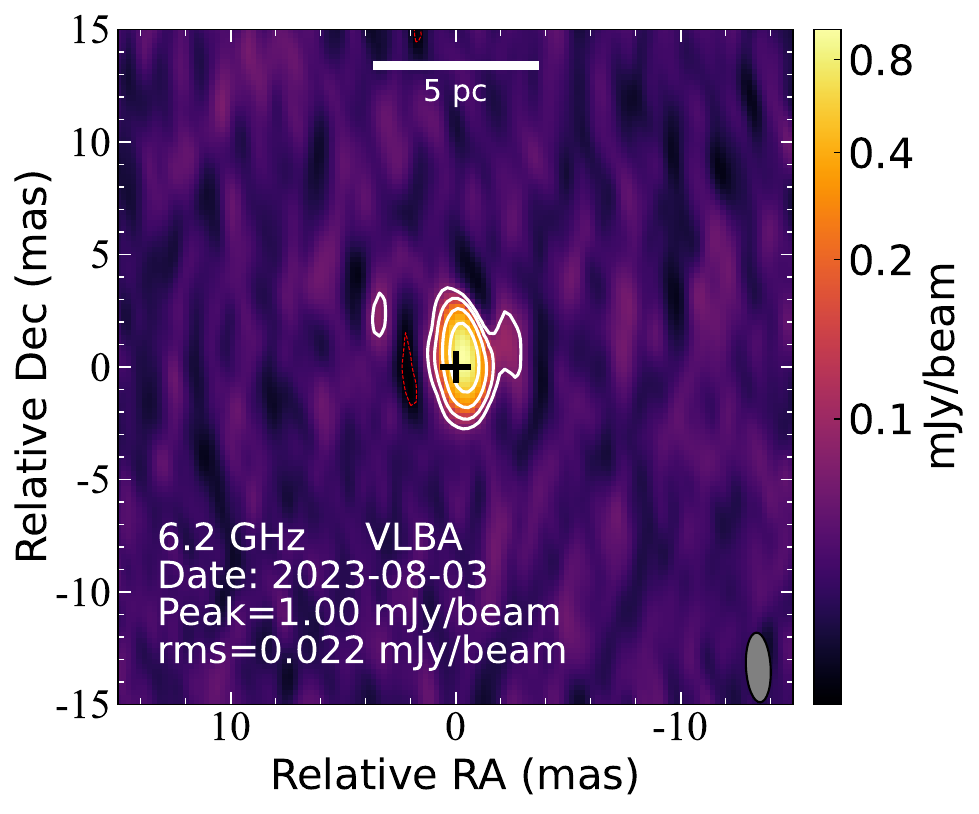}
    \includegraphics[width=0.25\linewidth]{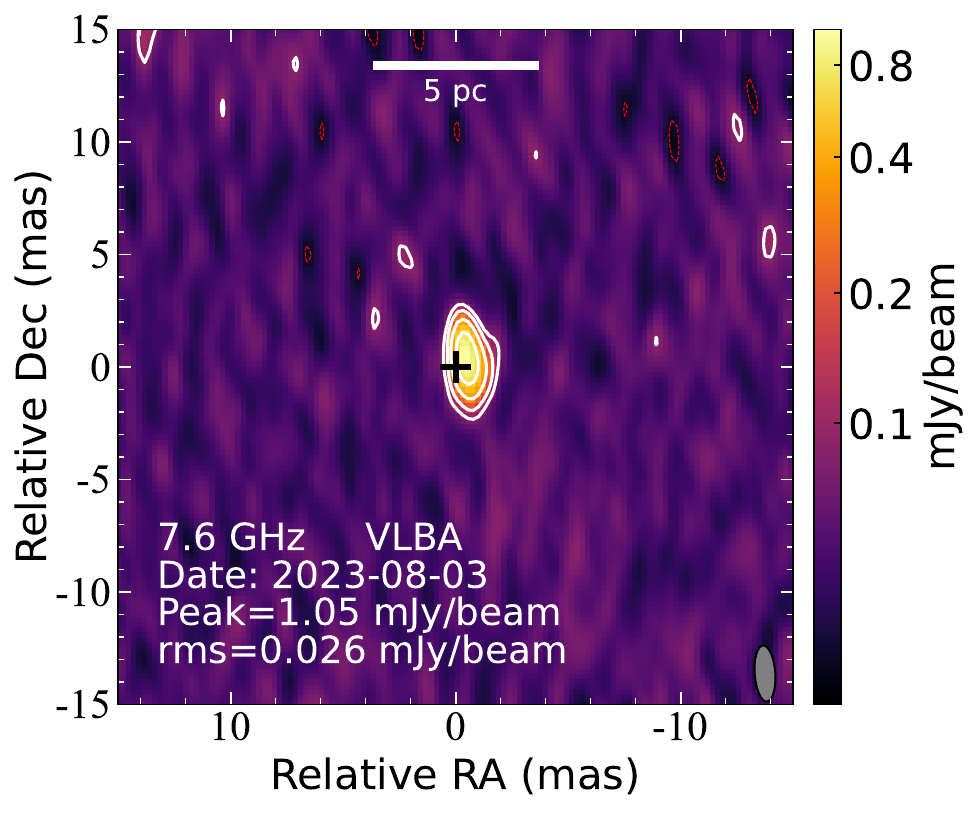}\\
    \includegraphics[width=0.25\linewidth]{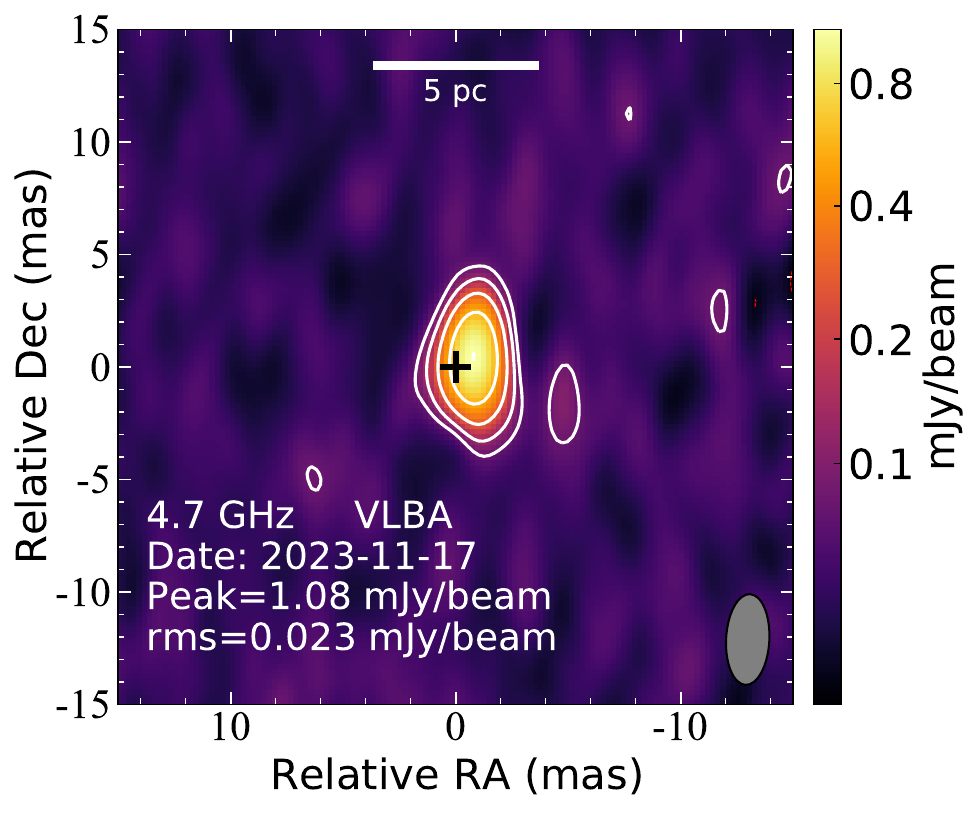}
    \includegraphics[width=0.25\linewidth]{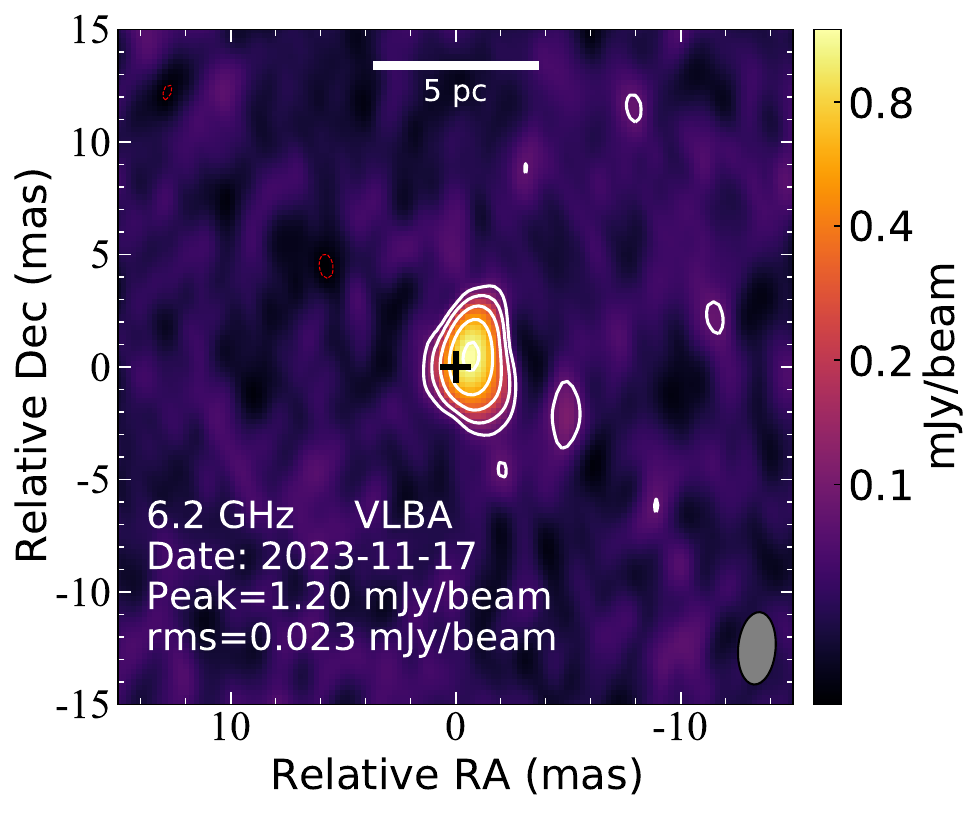}
    \includegraphics[width=0.25\linewidth]{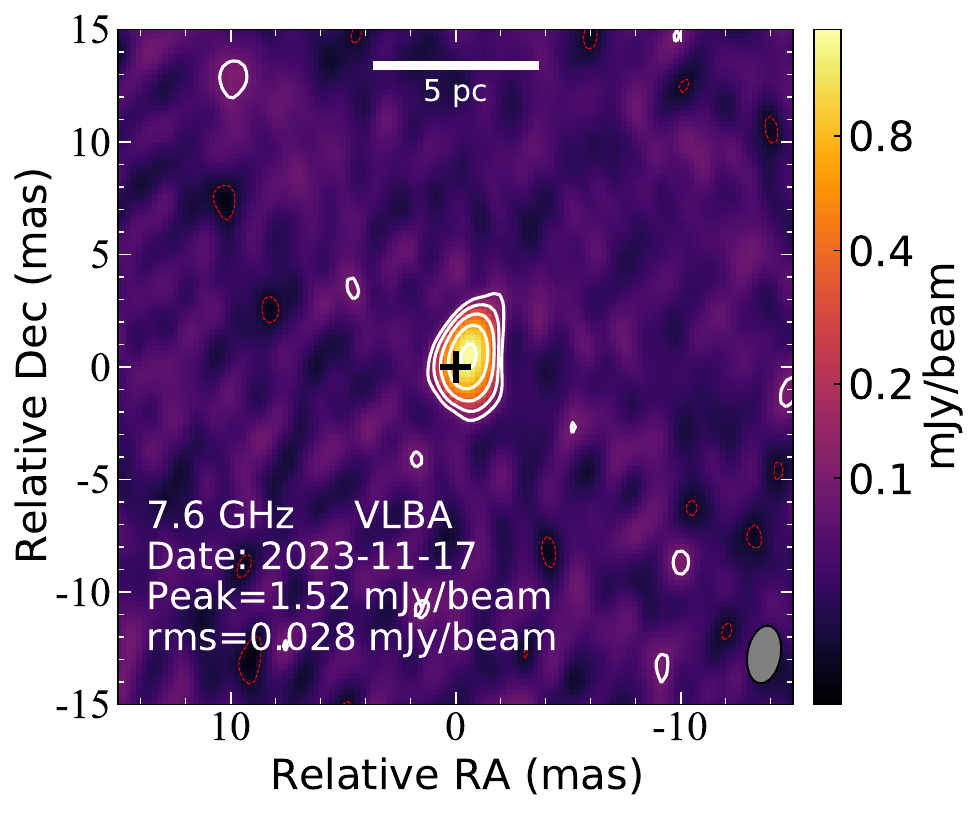}\\
    \includegraphics[width=0.25\linewidth]{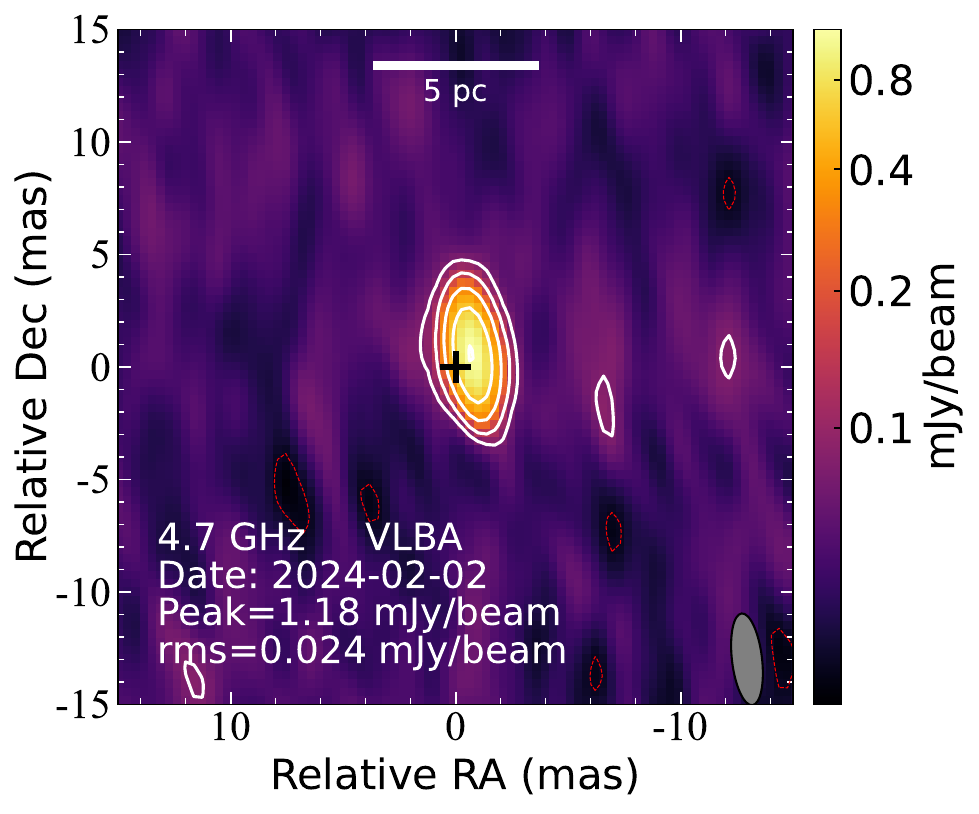}
    \includegraphics[width=0.25\linewidth]{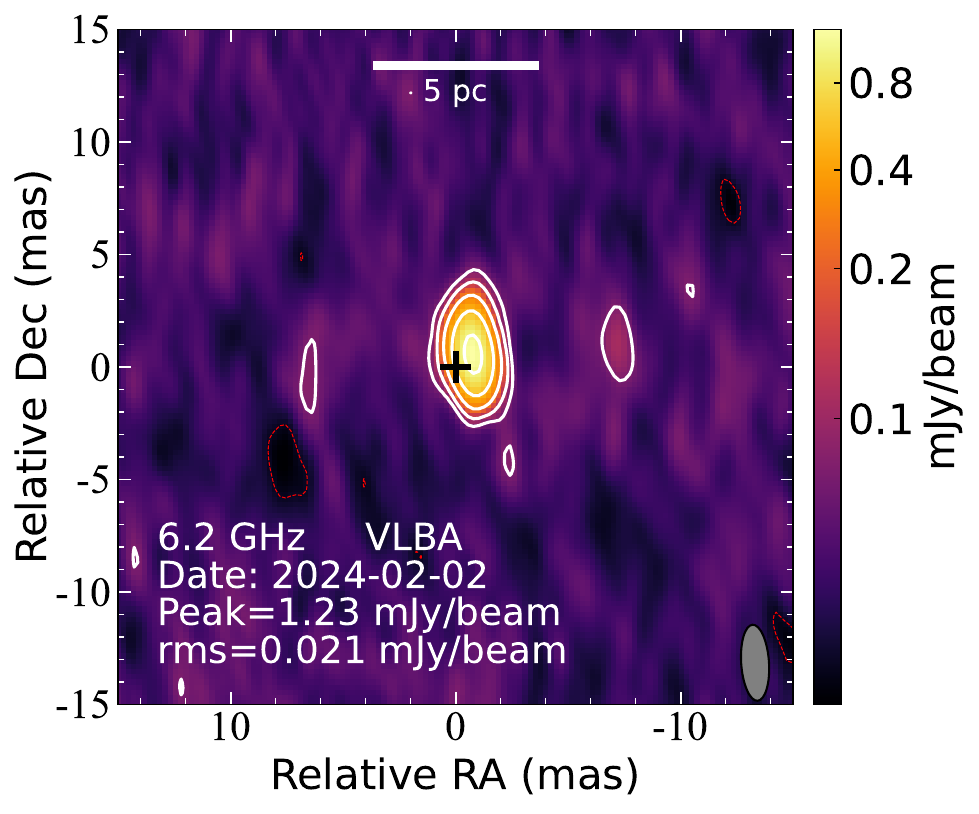}
    \includegraphics[width=0.25\linewidth]{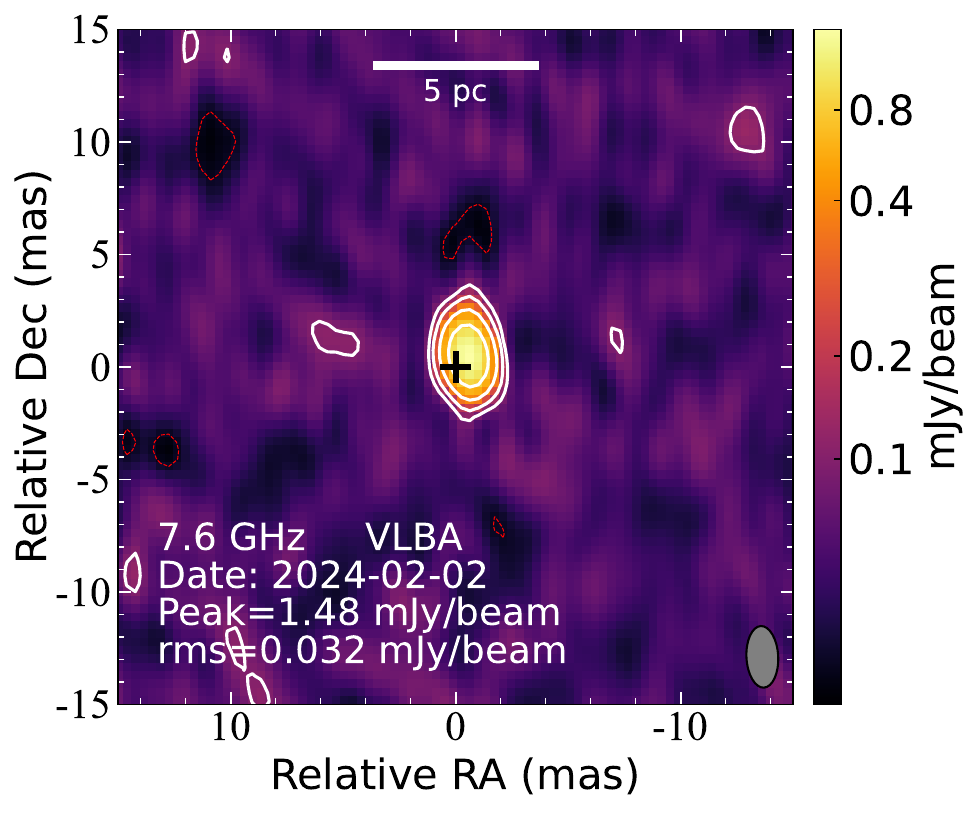}
    \caption{VLBI images of Mrk 110 are presented, with the observation date, peak intensity, and root mean square noise level indicated in the bottom-left corner of each map. The contour levels are 3 $\times$ rms $\times$ (1, 2, 4, 8, 16) mJy beam. The position of the \textit{Gaia} DR3 optical nucleus \citep{2022yCat.1355....0G} is marked by a cross at the center of each panel. The grey ellipse in the bottom-right corner illustrates the restoring beam's shape.}
    \label{fig:VLBI_image}
\end{figure*}

The VLBI observations spanning 2015--2024 reveal Mrk 110 as a compact radio source (Figures \ref{fig1} and \ref{fig:VLBI_image}). There is no significant extended emission detected above sensitivity thresholds (typically $3\sigma \sim 0.06$--$0.1$~mJy~beam$^{-1}$), placing stringent constraints on any potential jet-related extended emission, limiting it to $< 2.5$--$5\%$ of the core flux density.

We used the model-fitting program in the \texttt{Difmap} software package to quantitatively describe the emission structure. The source structure is best characterized by a single circular Gaussian component, as more complex models (e.g., core plus jet) do not provide statistically significant improvements in the fit. 

To rigorously evaluate whether a more complex model is warranted by our data, we conducted detailed model fitting using Markov Chain Monte Carlo (MCMC) methods. We compared two specific models: (1) a single circular Gaussian component and (2) an elliptical Gaussian component that could potentially represent an unresolved core-jet structure. For the MCMC analysis, we employed the emcee package with 200 walkers and 10,000 iterations after 5,000 burn-in steps. The parameter space for the circular Gaussian included position (x,y), flux density, and radius, while the elliptical Gaussian model additionally included major axis, ratio (major axis / minor axis), and position angle.
The posterior distributions from our MCMC analysis consistently demonstrated higher statistical significance for the single circular Gaussian model across all epochs. When forced to fit an elliptical Gaussian, the parameters failed to converge, further supporting the circular model.

Table \ref{tab:obs-info} lists the fitted parameters derived from the \texttt{Difmap} analysis (columns 6 and 7).  Detailed error estimates are provided in Appendix \ref{sec: error}. Using our highest resolution 4.9 GHz EVN observations from 2023 March 28, we derive a compact source size of 0.49$\pm$0.02 mas, corresponding to a projected linear size of approximately 0.34 pc. This extremely compact structure suggests efficient confinement of any potential outflow within the inner parsec of the AGN.

The extreme compactness of the Mrk 110 radio core, while similar to other radio-quiet NLS1s \citep{2013ApJ...765...69D, 2023MNRAS.518...39W}, shows notably higher flux densities and stronger pc-scale core dominance. This enhanced core dominance, combined with the later detection of relativistic jet features (see Section \ref{sec:3.1}), suggests Mrk 110 may represent a transition phase from relatively quiscent to active states.

\begin{figure*}
    \centering
    \includegraphics[width=0.5\linewidth]{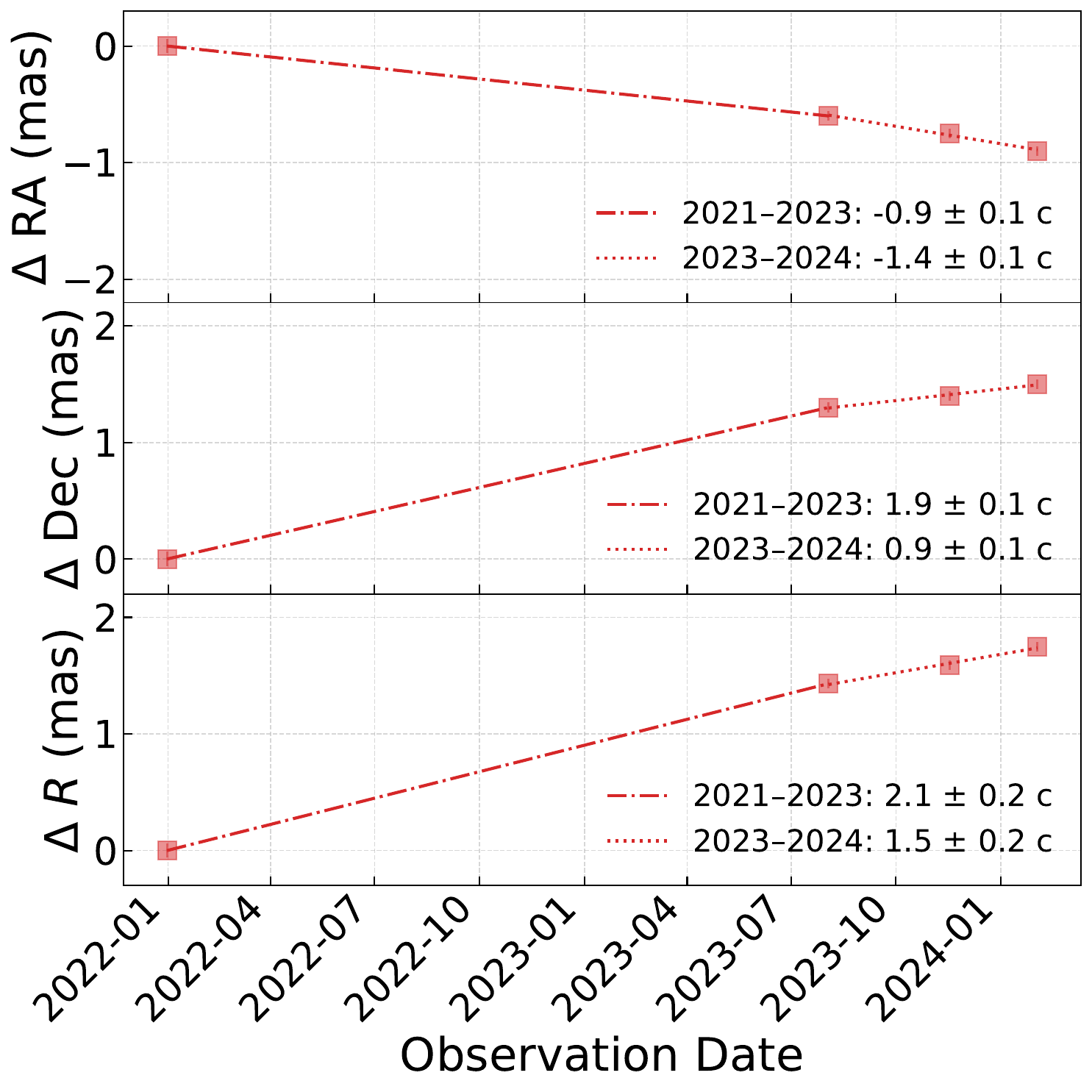}
    \caption{Proper motion of the radio core in Mrk 110 at 7.6 GHz derived from VLBA observations between 2021 and 2024. The top, middle, and bottom panels show the evolution of $\Delta$RA, $\Delta$Dec, and total displacement $\Delta$R (in mas), respectively, as a function of observation date. Two distinct motion phases are identified: an earlier phase from 2021 to 2023 (red dash-dot lines), and a later phase from 2023 to 2024 (red dotted lines). }
    \label{fig:pm_7GHz}
\end{figure*}

\section{Error estimate and analysis} \label{sec: error}

The uncertainty in VLBI flux density measurements is influenced by two main factors: a random error component, typically comparable to the image noise, and a systematic error associated with visibility amplitude calibration. For VLBA observations, this systematic error is generally 5\%, stemming from factors such as inaccuracies in antenna gain curves, opacity corrections, and other calibration parameters \citep{2009AJ....138.1874L}. For bright sources, systematic error often constitutes the dominant contribution to the total uncertainty.

The position uncertainty is calculated as the square root of the sum of squares of two components: (1) the statistical uncertainty, which is given by $\text{(beam size)} / (2 \times \mathrm{SNR})$, where $\mathrm{SNR} = S_{\rm peak} / \sigma_S$ is the signal-to-noise ratio of the fitted component, and (2) the positional error of the phase reference calibrator, which is typically 0.05~mas. Mathematically, the total position uncertainty is expressed as:
\[ 
\Delta \text{pos} = \sqrt{\left(\frac{\text{beam size}}{2 \times \mathrm{SNR}}\right)^2 + (0.05~\mathrm{mas})^2}
\].

\[
\Delta \text{pos} =\sqrt{ (\Delta \delta_{Ra})^2 + (\Delta \delta_{Dec})^2 }
\]

The error in the spectral index measurement originates from the uncertainties in the flux densities, which can be expressed using the equation:

\begin{equation}
\Delta\alpha =  \sqrt{\left(\frac{\Delta S_1}{S_1}\right)^2 + \left(\frac{\Delta S_2}{S_2}\right)^2}
\end{equation}

Here, $S_{1,2}$ represents the two flux density used to derive the spectral index.

Using the measured flux densities and sizes, we calculate the brightness temperature following \citet{1988gera.book.....K}:
\begin{equation}
    T_B = 1.22 \times 10^{12} \frac{S}{\text{mJy}} \left( \frac{\nu}{\text{GHz}} \right)^{-2} \left( \frac{\theta}{\text{mas}} \right)^{-2} (1 + z) \, \text{K},
\end{equation}
where $S$ is the flux density, $\nu$ is the observing frequency, $\theta$ is the angular size, and $z$ is the redshift. We estimate the error of the calculated brightness temperature using error propagation:
\begin{equation}
\Delta T_B = T_B \times \sqrt{\left(\frac{\Delta S}{S}\right)^2 + 4\left(\frac{\Delta \theta}{\theta}\right)^2 }    
\end{equation}
Where $\Delta S$ is the uncertainty in flux density, and $\Delta \theta$ the uncertainty in angular size.

\section{International LOFAR observations, data processing and image} \label{app:lofar}

The LOw-Frequency ARray (LOFAR) High Band Antennas (HBAs) observation of Mrk\,110, presented in this paper, was conducted on 2016-08-16 (Project code: LC6\_015). The phase centre of the observation was directed at J2000 RA = $09^\mathrm{h}32^\mathrm{m}6\fs48$, Dec = $52\arcdeg03\arcmin32\farcs4$, with the target observed over a standard 8-hour period. This observation provided a maximum projected baseline of 1468~km, resulting in a synthesized beam size of $0\farcs33 \times 0\farcs23, 179\degr$ at 144~MHz. Data were initially recorded at a 1-second cadence and 3~kHz resolution, then averaged to 1 second and 12~kHz resolution before being uploaded to the long-term archive. The total bandwidth of 48~MHz covered the frequency range from 120 to 168~MHz. Observations of two compact flux density calibrators, 3C\,196 and 3C\,295, bracketed the target observation of P143+52. Each calibrator scan lasted 10 minutes and used the same configuration as the target observations. A detailed description of the observing procedure and the data processing pipeline can be found in \cite{2022A&A...658A...1M}.

The international LOFAR 144 MHz image of Mrk 110 reveals a kpc-scale two-component structure (Fig. \ref{fig1}), with the brightest component centered at RA = 09:25:12.9077, Dec = 52:17:10.0958 (peak flux: 3.03 $\pm$ 0.13 mJy/beam; integrated flux: 10.51 $\pm$ 0.58 mJy) and the northern component at RA = 09:25:12.9684, Dec = 52:17:12.4170 (peak flux: 0.58 $\pm$ 0.05 mJy/beam; integrated flux: 6.30 $\pm$ 0.55 mJy). The international LOFAR results for Mrk 110 are consistent with the findings of early VLA observation \citep{1994AJ....108.1163K, 1998MNRAS.297..366K}. 
The brightest lobe’s location is consistent to the compact nuclear region detected by VLBI, indicating a physical connection between this component and the currently active jet core.
The brightest lobe’s low peak-to-integrated flux ratio ($S_{peak}/S_{int}=0.28$) indicates synchrotron cooling dominance in this region, where the electron energy spectrum has aged significantly due to radiative losses, while the northern lobe’s elongated morphology and extremely low ratio ($S_{peak}/S_{int}=0.09$) suggest synchrotron cooling dominance in this region, where the electron energy spectrum has aged significantly due to radiative losses. 
The extended morphology of the northern lobe may result from anisotropic interactions between the AGN jet and the galactic medium, such as the propagation of shock waves or constraints imposed by magnetic field alignment \citep[e.g., ][]{2007MNRAS.376.1849H}.
The lobes are separated by $2.4\arcsec$ (1.7 kpc) north, corresponding to a dynamical timescale of $(0.5-5.4) \times 10^5$ yr for jet velocities of $0.01-0.1\, c$. Their alignment with the VLBI-scale relativistic jet, demonstrates stable jet orientation over Myr timescales. 

\section{Quantitative Comparison with Theoretical MAD Models}

The observational data presented in the main text provide strong evidence for relativistic jet formation in Mrk 110, a traditionally radio-quiet AGN. Here, we quantitatively evaluate these findings within the theoretical framework of magnetically arrested disk (MAD) models to establish a more robust connection between our observations and physical jet-launching mechanisms.

Recent GRMHD simulations of MAD states predict specific jet parameters that closely match our observations of Mrk 110. Quantitatively, our measured Lorentz factor ($\Gamma_{\min} = 2.3$) falls within the range predicted by \citet{2023arXiv231100432C} for transient MAD states in systems with similar black hole mass and accretion rate (predicted $\Gamma \approx 2$--$5$). 
Although the Eddington ratio of Mrk 110 ($\dot{m} \approx 0.40$) is relatively high compared to typical low-luminosity MAD systems, the predicted jet properties still qualitatively match.

The observed episodic behavior corresponds to the MAD formation/dissipation cycles predicted for a black hole of mass $\sim 2\times10^7 M_{\odot}$, where magnetic flux saturation occurs on timescales of $t_{\rm MAD} \approx 10^3 - 10^5~r_{\rm g}/c$  \citep{2011MNRAS.418L..79T, 2023arXiv231100432C}. While current simulations capture shorter-timescale MAD processes, the observed year-level time span likely represents a higher-order cycle involving global disk reconfigurations and interactions with the surrounding galactic environment that exceed typical simulation durations. This correlation between the observed recurrence timescale and theoretical predictions provides support for the MAD interpretation.

We can estimate the dimensionless magnetic flux parameter $\phi_{\rm BH}$ for Mrk 110 using our observed jet power and the relation from \citet{2011MNRAS.418L..79T}:

\begin{equation}
\eta_{\rm jet} \approx \frac{P_{\rm jet}}{\dot{M}c^2} \approx \kappa \phi_{\rm BH}^2 a_*^2
\end{equation}

where $\kappa \approx 0.05$ is an efficiency factor, $a_* \approx 1$ is the dimensionless black hole spin \citep{2024A&A...681A..40P}, and $\eta_{\rm jet}$ is the jet production efficiency. Using our measured jet power ($P_{\rm jet} \approx 7.37\times10^{42}$ erg s$^{-1}$ ) and the estimated accretion rate ($\dot{M} \approx 0.17~M_{\odot}$ yr$^{-1}$, corresponding to $\dot{m} \approx 0.40$ for a $2\times10^7 M_{\odot}$ black hole), we obtain $\eta_{\rm jet}$ $\approx$ 7.3 $\times 10^{-4}$ and $\phi_{\rm BH} \approx 0.12$, below the MAD threshold ($\phi_{\rm MAD} \approx 50$) for steady-state jet production but consistent with transient MAD episodes.
This derived value of $\phi_{\rm BH}$ places Mrk 110 in the regime where episodic MAD states can form temporarily but are not sustained continuously, explaining the intermittent jet launching we observe. This quantitatively supports our interpretation that radio-quiet AGN like Mrk 110 can occasionally satisfy jet-launching conditions through temporary magnetic flux accumulation.

The pronounced jet deceleration we observe at approximately 3--6 pc (deprojected) from the core coincides with the transition zone between the broad-line region (BLR) and narrow-line region (NLR), consistent with theoretical predictions of jet-ISM interactions. Recent simulations by \citet{2024MNRAS.532.1522J} predict that MAD jets interacting with dense environments characteristic of high-Eddington ratio AGN should experience significant deceleration at precise distances where the external medium pressure becomes comparable to the jet pressure. For Mrk 110's specific parameters, this transition is expected at a distance beyond $10^{18}$ cm, roughly consistent with the observed deceleration distance of 3--6 pc.

The spectral evolution we observe—from steep ($\alpha \approx -0.63$) to inverted ($\alpha \approx +0.69$)—also quantitatively matches theoretical predictions for emerging self-absorbed synchrotron jets. The initial spectral index of $\alpha \approx -0.63$ is consistent with the canonical value for optically thin synchrotron emission from a homogeneous source with a power-law distribution of relativistic electrons with energy index $p \approx 2.4$ (yielding $\alpha = 5/2 - (p+4)/2 \approx -0.7$; \citealt{1985ApJ...298..114M}).

Furthermore, the observed variation in jet speed between episodic events remains within a relatively narrow range ($\beta_{\rm app} \approx 2.1$--$3.7$), suggesting a stable jet production mechanism with consistent energetics. This stability in jet parameters across multiple ejection episodes, separated by years, strongly supports a recurring physical process rather than stochastic variability, further strengthening the connection to cyclic MAD state formation.

In combination, these quantitative comparisons between our observational results and theoretical predictions from MAD models provide compelling evidence that episodic jet launching in Mrk 110 is driven by the temporary establishment of MAD conditions in an otherwise sub-MAD accretion flow. This unified framework successfully explains the observed jet kinematics, spectral evolution, and recurrence pattern, while placing Mrk 110 within the broader context of AGN jet formation physics.

\section{Estimation of jet kinematic parameters}

The observed positional shift of the radio core suggests a minimum apparent velocity of $\beta_{\rm app} > 2.1$, under the assumption that the jet ejection occurred in early 2022. Applying the standard relativistic beaming framework, this apparent speed constrains the jet kinematic parameters. Specifically, it implies a minimum Lorentz factor of $\Gamma_{\rm min} \geq 2.3$ when cos $\theta$ = $\beta$, following the relation $\Gamma_{\rm min} = \sqrt{1 + \beta_{\rm app}^2}$, and a corresponding maximum viewing angle of $\theta \leq 25.5^\circ$ \citep{1993ApJ...407...65G}. 

In the later phase of the jet evolution, the positional measurements correspond to an apparent velocity of $\beta_{\text{app}} = 1.5 \pm 0.2$. Using the same relativistic beaming theory, this lower apparent speed yields updated constraints: a minimum Lorentz factor of $\Gamma_{\text{min}} = 1.8$, and a maximum viewing angle of $\theta \leq 33.7^\circ$. These calculations illustrate a deceleration of the jet over time, transitioning from a highly relativistic initial ejection phase to a mildly relativistic flow, which is a common feature in AGN jets interacting with their surrounding environment.

\end{document}